\documentclass[aps,pre,
 amsmath,amssymb]{revtex4-1}
\usepackage{graphicx}% Include figure files
\usepackage{bm}% bold math
\usepackage[usenames, dvipsnames]{color}

\usepackage[utf8]{inputenc}
\usepackage[T1]{fontenc}
\usepackage{mathptmx}
\usepackage{hyperref}
\usepackage{subcaption}
\usepackage{caption,setspace}
\usepackage{multirow}

\renewcommand{\thefigure}{ Fig. \arabic{figure}}

\usepackage[toc,title,page]{appendix}

\begin{document}

\title{A dual-space classification scheme of spatial heterogeneities in 2D monatomic supercooled liquids}
\author{Viet Nguyen}
\author{Xueyu Song}
\email{xsong@iastate.edu, vnguyen@iastate.edu}
\affiliation{Ames Laboratory and Department of Chemistry, Iowa State University, Ames, IA, USA}
\date{\today}

\begin{abstract}

Understanding the physics of supercooled liquids near glassy transition remains one of the major challenges in condensed matter science. There has been long recognized that supercooled liquids have spatially dynamical heterogeneity ~\cite{Cubuk,Walter_Kob,Sastry,M.D.Ediger} whose dynamics in some regions of the sample could potentially be orders of magnitude faster than the dynamics in other regions only a few nanometers away. However, to identify such domain structures both structurally and configurationally in a consistent fashion and the connection between structures and dynamics remains elusive. A new approach to classify these spatial heterogeneities of supercooled liquids is developed. Using average weighted coordination numbers (WCNs) of particles as features for the Principle Component Analysis (PCA) and Gaussian Mixture (GM) clustering, a representation of the feature space to perform GM clustering is constructed after applying PCA to WCNs. Nano-domains or aggregated clusters are found in configurational (real) space by direct mapping the identities of clusters in the feature (structural) space with some classification uncertainties. These uncertainties can be improved by a co-learning strategy to transmit information between the structural space and configurational space iteratively until convergence. The domains are indeed found to be remarkably consistent in both spaces over long times as well as measured to have heterogeneous dynamics. 
Using such a classification scheme the order parameter time correlation function follows the non-conserved order parameter scaling law, which indicates that such domains are the result of liquid-liquid phase separation after quenching.
\end{abstract}
\maketitle

\section{Introduction}

Phenomena of spatial heterogeneity in supercooled liquids near glass transition have been studied intensively and widely accepted over past few decades due to considerable experimental ~\cite{E.VidalRussell,WKKegel,Marcus} and theoretical ~\cite{Walter_Kob,DonnaN.Perera,B.Doliwa} efforts. Observation of heterogeneous dynamics is closely related to the plateau  region of the relaxation dynamics and the onset of caging process, where particles with the similar  mobility tend to move in cooperatively manner that form dynamically correlated mesoscopic domains or clusters.~\cite{C.PatrickRoyall,AndreaCavagna}. Several theories of glass transition have been developed for a fundamental description of these spatial domains such as the energy landscape picture~\cite{Goldstein_1969}, the Adam-Gibbs theory~\cite{Adam_Gibbs_1965} and the random first-order transition theory~\cite{RFOT}. These theories are linked explicitly to thermodynamics of the system to characterize the formation of domain structures. Despite of their simple and intuitive description to the spatial domains of liquids, it lacks 
reliable methods to characterize them. Meanwhile, some classification approaches like Voronoi polyhedra, bond-orientational order parameters ~\cite{Boattini:2020vp,Spellings,Emanuele}, the common-neighbour analysis ~\cite{Reinhart}, and topological clusters require some specific structures $\it a$ $\it priori$ of the system which is unknown in general while some other  ‘order-agnostic’’ approaches utilize mutual information based on Shannon entropy lacks microscopic structural details of the system~\cite{C.PatrickRoyall,Royall_2015}. 

In this study, a new classification approach is developed to identify these spatially heterogeneous domains showing improvement over previous methods. In our approach, radial distribution function $\it g(r)$ does not play a role in qualitatively delivering signatures of structural heterogeneities~\cite{Sastry} but becomes an order-parameter to unlock the picture of heterogeneous domains by a collection of shell coordination numbers (CNs). The collection of CNs is highlighted by two points: collecting from a simulation frame to frame to avoid the statistically average of structures and employing a Gaussian weight to overcome a conventional cut-off distance criterion to scale the contribution of neighboring particles continuously within a shell. With aids of some machine learning (ML) algorithms such as Principle Component Analysis (PCA), K-means clustering and Gaussian Mixture (GM), the information of CNs are fully extracted out to reveal the formation of nano-domains both in structural and configurational spaces consistently. These nano-domains from our classification scheme indeed reflect widely accepted spatial heterogeneities phenomena. 

By taking advantage of the classification scheme, nature of these domains is clearly clarified and the spatially correlated heterogeneous dynamics are characterized: the intra-domain relaxation is primarily due to the same mobility of particles inside the domain and inter-domain relaxation which influences the morphological evolution of nano-domains structures. Furthermore, the slow inter-domain relaxation is related to the coarsening kinetics of liquid-liquid phase separation. 

In this paper, the method is implemented for a 2D simple monatomic system with Lennard-Jones Gauss (LJG) interatomic potential~\cite{2D_Mizuguchi,MichaelEngel,Hoang:2011wa}. The reason to select this 2D model is because this LJG potential leads to glass transition which normally occurs with more complex systems~\cite{2D_Mizuguchi}  and the formation of nano-domains in 2D can be visualized easily to provide a clearer picture of the phenomenon than in 3D~\cite{Nguyen_2022, Nguyen_2023}. 

The paper is organized as follows. In Sec. \ref{sec:methods}, our MD simulated model and proposed methods are described in detail. This is followed by an extensive results presentation with discussions in Sec. \ref{sec:results}. Some concluding remarks are given in Sec. \ref{sec:conclusion}.

\section{Methodology} \label{sec:methods}

\subsection{Simulation Details}

\begin{figure*}[!htb!]\centering

\includegraphics[width=0.45\columnwidth]{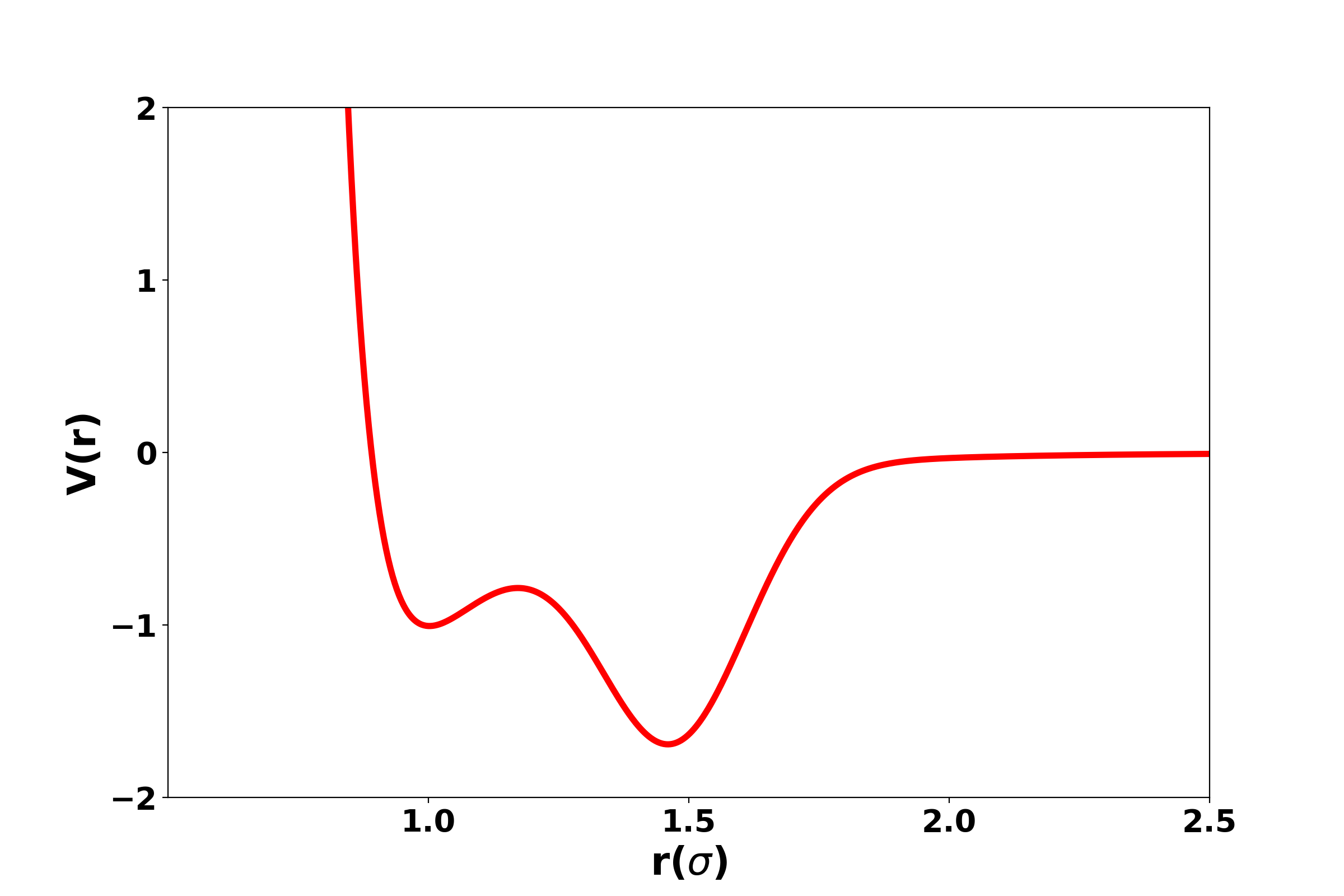}
\caption{The plot of LJG potential as a function of distance between two atoms.} 
\label{fig:potential}
\end{figure*}

We consider a system of particles which interact isotropically through the LJG potential whose form is given by: 
\begin{equation}
V(r)= \epsilon \left[ \left( \frac{\sigma}{r} \right)^{12} - 2\left( \frac{\sigma}{r} \right)^6 \right] - 1.5\epsilon \left[  \frac{\left(r - 1.47\sigma\right)^{2}}{0.04\sigma^2} \right],
\end{equation} where the parameter $\epsilon$ is the potential well depth and $\sigma$ is the characteristic atomic diameter. The LJG potential is a double-well potential where the first well is the LJ potential and the second well is represented by a Gaussian function as shown in \ref{fig:potential}.  The second minimum is added at a typical range of second-nearest neighbor coordination distance in close-packed crystals, which suppress the crystallization of the system. The values of parameters for the position, depth and width of the Gaussian part are selected so that glassy states are formed. It was found that the existence of the second well with chosen parameters favors the formation of a pentagonal local order and pentagons are essential for stability of a glassy state~\cite{2D_Mizuguchi}. Hence, the LJG potential is a suitable model for our study of  simple supercooled and glass-forming liquids. 

All MD simulations of the system are performed using LAMMPS ~\cite{LAMMPS} with $\it{NVT}$ ensemble. The cut-off distance is set to 2.5$\sigma$ and periodic boundary conditions are applied to all directions. All units are in LJ-reduced units in line with parameters of solid Ar~\cite{Vega}:  the time unit in term of $\tau=t\sqrt{m\sigma^2\over\epsilon}$ and reduced temperature is defined by $T^* = T(\frac{\epsilon}{k_B})$where $k_B$ is the Boltzmann constant. The Verlet algorithm is employed and MD time step is $0.005\tau$ or about 10 fs for Ar parameters. Initial atomic configuration of 2D simple square lattice of 6400 particles with a fixed density $\rho (N/A) = 1.0$  is heated up to a high temperature to obtain a liquid state and relaxed for 2 x $10^6$ time steps. The system is then quenched to a target temperature $T^*$ = 0.35 or 0.2,  and is equilibrated again before collecting configurations at every 100 time steps or $0.05\tau$ in the production run. The cooling process has been done by linearly decreasing temperature via re-scaling atomic velocity: $T = T_0 -\gamma n$, where $\gamma$  is the cooling rate (9.67 x $10^{10}$ K/s if taking Ar parameters) and $n$ is the number of MD steps. The cooling rate we used is similar to the cooling rate in the reference~\cite{Spellings}. At this cooling rate which is effectively a quenching, the model shows a metastable supercooled liquid state for a long time, while the crystallization is suppressed. For each temperature, three independent runs are generated to improve statistics and consistency of the results. These temperatures are selected because: $T^*$ =  0.35 is just above the glass transition temperature ($T^*_g = 0.31$ ) predicted by ~\cite{VoVanHoang}  to be able to observe substantial change of dynamics while $T^*$ = $0.2$ is below the $T^*_g$ to study the trend of structural heterogeneity for temperature dependence.

\subsection{Radial Distribution Function (rdf) and Weighted Coordination Numbers (WCNs)} \label{sec:rdf}

\begin{figure}[h!] 
\resizebox{\columnwidth}{!}
{
\begin{subfigure}{0.35\textwidth}
\includegraphics[width=2.1in]{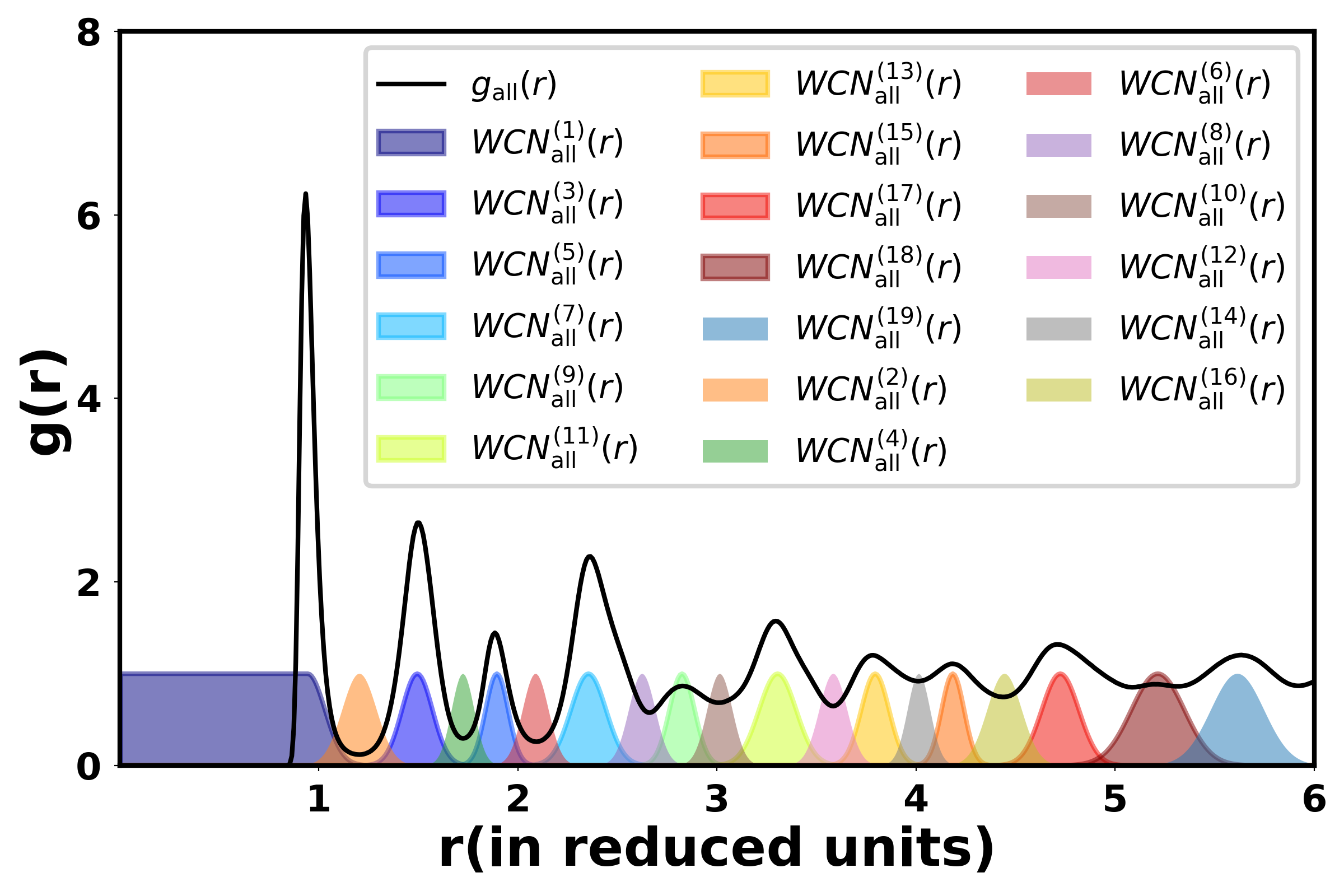}
\caption{WCNs based on rdf}
   \label{fig:rdf}
\end{subfigure}
\hfill
\begin{subfigure}{0.35\textwidth}
\includegraphics[width=2.1in]{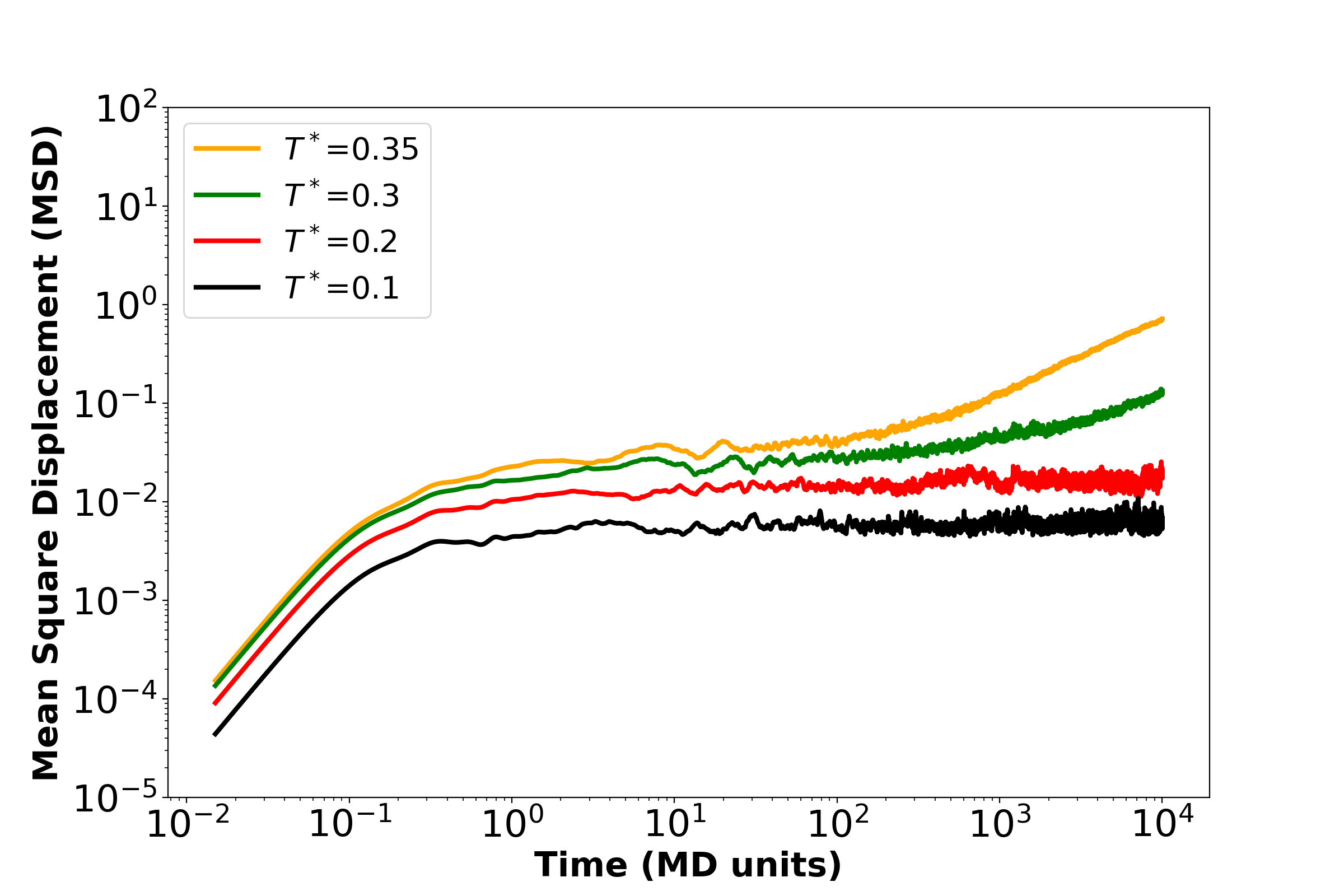}
\caption{MSD trajectories}
   \label{fig:msd}
\end{subfigure}
}
%\parbox{3.3in}
%{
\caption{ (a) Radial distribution function (rdf) employed by weighted Gaussian functions for the system at $T^*=0.2$. The color and superscript indicate the WCN's position and number along the rdf.  b) All-particle MSDs as a function of time at various temperatures}  
\label{fig:decomposition}
%}
\end{figure}

To detect hidden heterogeneous structures of a disordered system, order parameters are required to accurately characterize its spatial local environment. In one extreme, the radial distribution function {\it g(r)} has been commonly used to describe the normal liquid structure by means of aggregating coordination numbers (CNs) in all solvation shells, {\it i.e.}, the relative number of neighboring particles in a particular surrounding spherical shell averaging over many configurations of all particles, hence a highly averaged function that lacks the details to capture the heterogeneous structural features of a super-cooled liquid. At another extreme, for a particular configuration of the system either a snapshot of simulations or an experimental image, local structures of an {\textbf  M}  particles system can be given by {\textbf  M}  particles' local coordination shell structure. Naturally,  a middle ground for a useful structural description of the system will be to classify these local structures into a few meso-states, which represent the overall structural heterogeneity of the system. Then the particles in the same meso-state should form domains in the configurational space, and these domains from different meso-states together can tile up the whole system, which provide a classification scheme both structurally and configurationally. Furthermore, lifetime of the meso-states in the structural space and domains in the configurational space is in the order of nano-second timescale which is long enough to make further analysis possible. The formation of meso-states (PC-space) or nano-domains (real space) are attributed to the plateau region or caging process. In the \ref{fig:msd}, the plateau region in diffusion at various temperatures shows the salient features of the onset of caging processes observed in supercooled liquids near glass transition~\cite{kob_andersen_PhysRevE.51.4626} and ranges from 5x$10^1$ to 2x$10^4$ MD units or converted to 0.1 to 40 (ns) for Ar parameters ~\cite{Schroder}. Thus, the lifetime of nano-domains is indeed  longer than or at the order of nanoseconds in the supercooled liquids.

From a snapshot of MD simulations, the CNs of each particle in that configuration can be calculated. However, the major drawback of conventional CN-based features is the discontinuity of counting neighboring particles to be in or out of a particular shell. To alleviate this strict assignment, weighted coordination numbers (WCNs) ~\cite{rudzinski_1.5064808}, which utilize normalized Gaussian distributions on the shell structure (\ref{fig:rdf}), are employed to weight the contribution of each surrounding particle based on the particle's distance to the central  one. WCNs are proved to be a better choice because it could count the fractional contribution of  particles based on the Gaussian distribution in each shell, hence it smoothes out the discontinuity among solvation shells. After identifying relevant solvation shells at various maxima and minima along its rdf, normalized Gaussian weighting functions are placed at the center of those shells. The width of the Gaussian function is determined by an area of associated solvation feature and intersecting tolerance of 0 to 0.25 with neighboring Gaussians is chosen in this study, but other reasonable choices of width and the size of the overlapping areas of the Gaussians do not change the consistency of the final results. The dimension of WCNs vector in each configuration is equivalent to number of shells in the {\it g(r)}, N = 19 is shown in the \ref{fig:rdf}, where the value of each WCN is the summation of all weighted surrounding particles within that shell in a single configuration; other reasonable numbers of shells tested yield the same results. For each configuration, employing this WCNs implementation, the features data for all particles in a particular configuration is represented by a matrix ${\widetilde{\bf X}}$ of {\textbf {M{\rm x}N}},   which consists of {\textbf N} WCNs for each of the {\textbf  M} particles. Such collected WCNs are still noisy in a disordered system, hence a smoothing is achieved by using an average of WCNs: $\overline{WCN}_i=\frac{1}{N_b}{\sum_{j}^{N_b}WCN_j}$, where $N_b$ is the number of neighboring particles in each shell plus the particle $i$ itself. Each particle's  {\textbf  N}  features in the $\overline{WCNs}$ matrix is constructed independently to describe its own local environment with respect to its surrounding particles, thus doesn't correlate to each other to provide order parameters  that reveal the hidden structures.

\begin{figure}
\resizebox{\columnwidth}{!}
{
\begin{subfigure}{0.35\textwidth}
\includegraphics[width=1.8in]{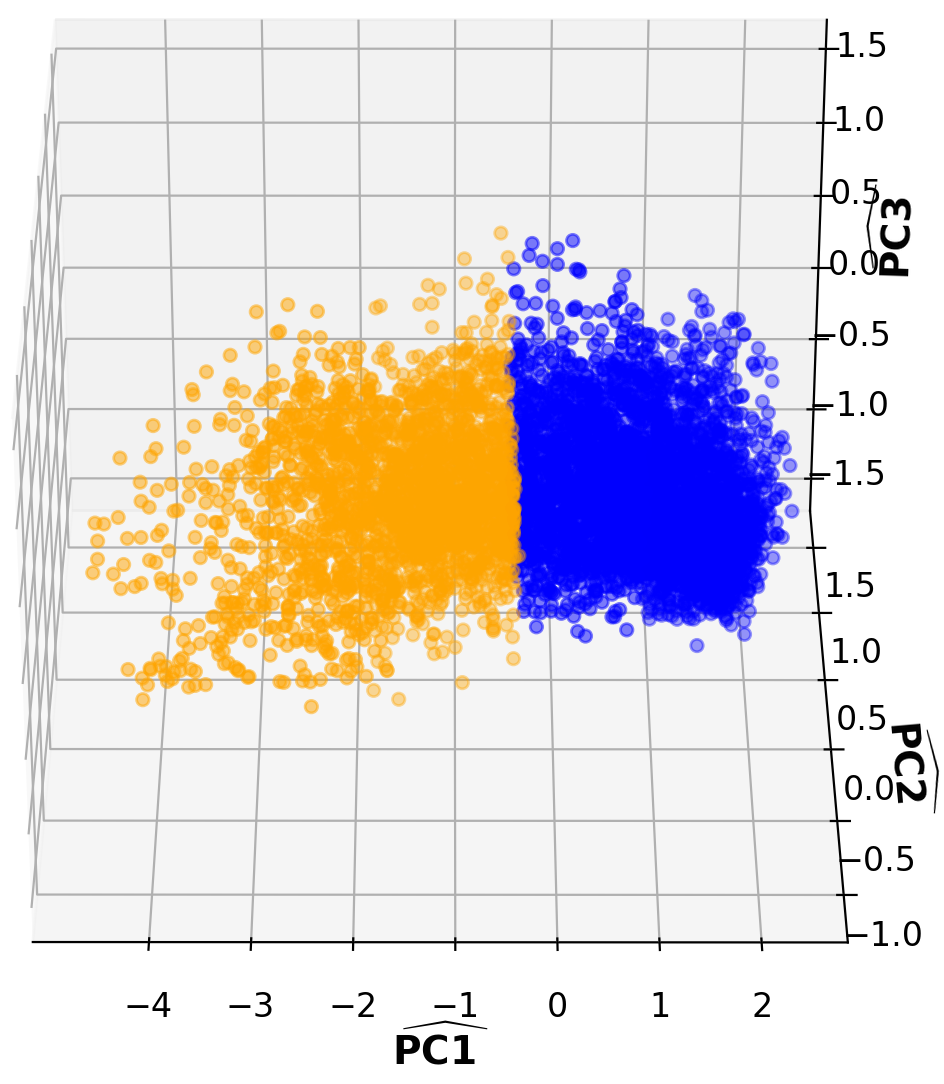}
\caption{}
   \label{fig:a}
\end{subfigure}
\hfill
\begin{subfigure}{0.35\textwidth}
\includegraphics[width=1.8in]{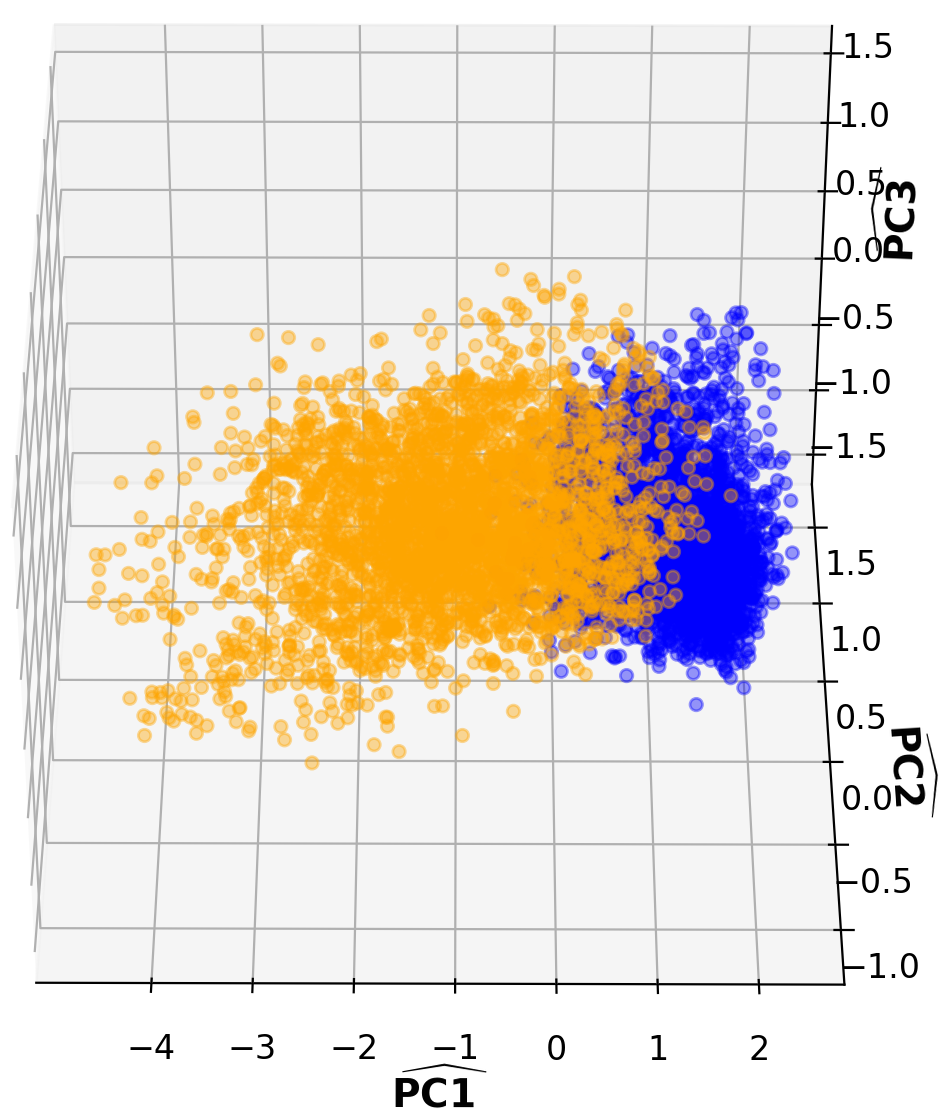}
\caption{}
   \label{fig:b}
\end{subfigure}
\hfill
}
\resizebox{\columnwidth}{!}
{
\begin{subfigure}{0.35\textwidth}
\includegraphics[width=1.9in]{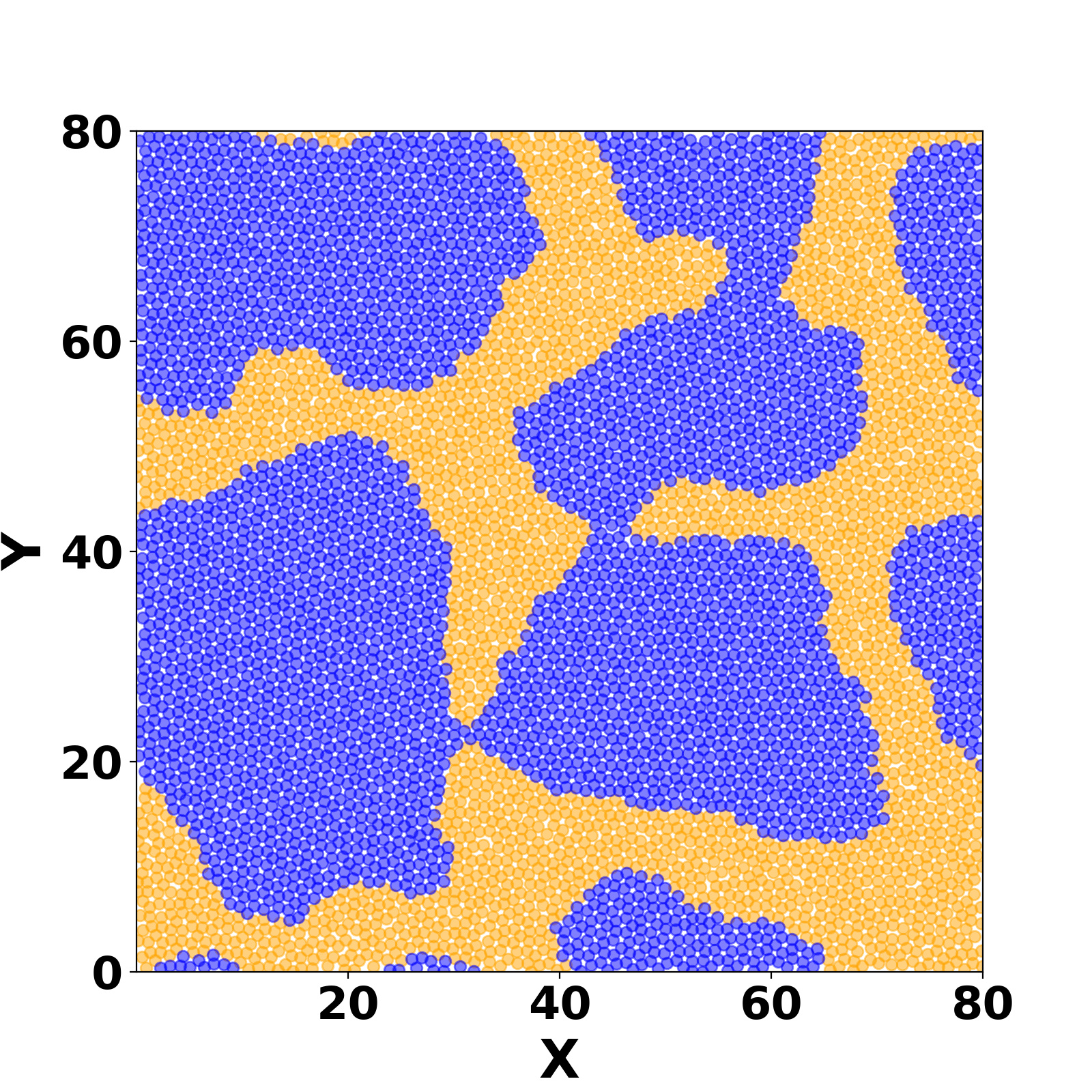}
\caption{}
   \label{fig:c}
\end{subfigure}
\hfill
\begin{subfigure}{0.35\textwidth}
\includegraphics[width=1.9in]{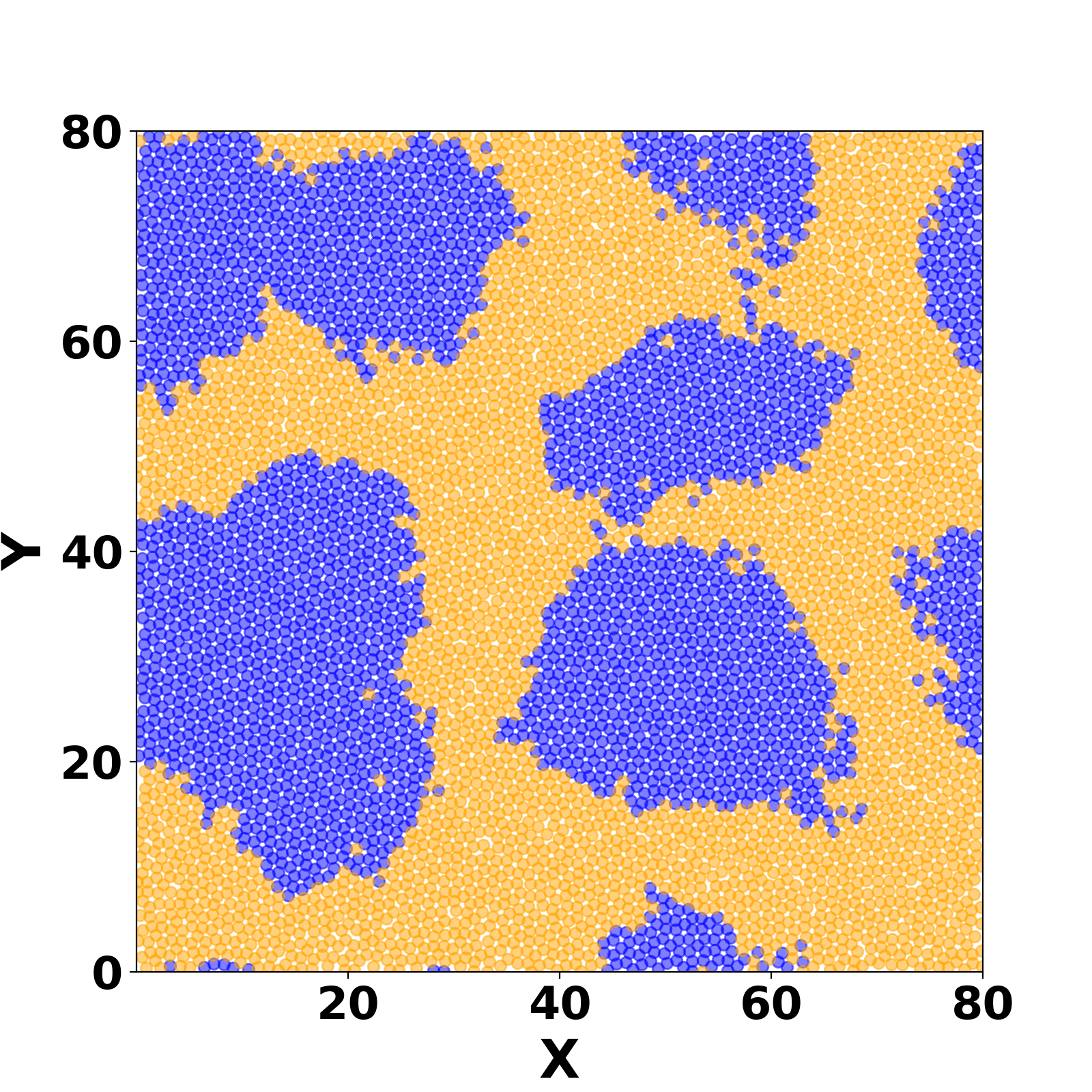}
\caption{}
   \label{fig:d}
\end{subfigure}
}
%\parbox{3.3in}
%{
\caption{ A 2D snapshot of 2 meso-states of 6400 particles system at $T^*=0.35$, for both direct mapping and co-learning method. Panels a) and c) are for the PC-space and configurational space from direct mapping while panels b) and d) present clusters formed in the PC-space and configurational space after convergence from co-learning. The hat symbol for labelling axes of the PC space represents the inner product of a particle's $\overline{WCN}$s with the PCs basis, in this case for the first three PC components. Blue, orange-colored particles corresponds to meso-state 1 and meso-state 2, respectively. X,Y are the atomic coordinates in reduced units.}  
\label{fig:iteration}
%}
\end{figure}

\begin{figure}[h!] 
\resizebox{\columnwidth}{!}
{
\begin{subfigure}{0.35\textwidth}
\includegraphics[width=1.8in]{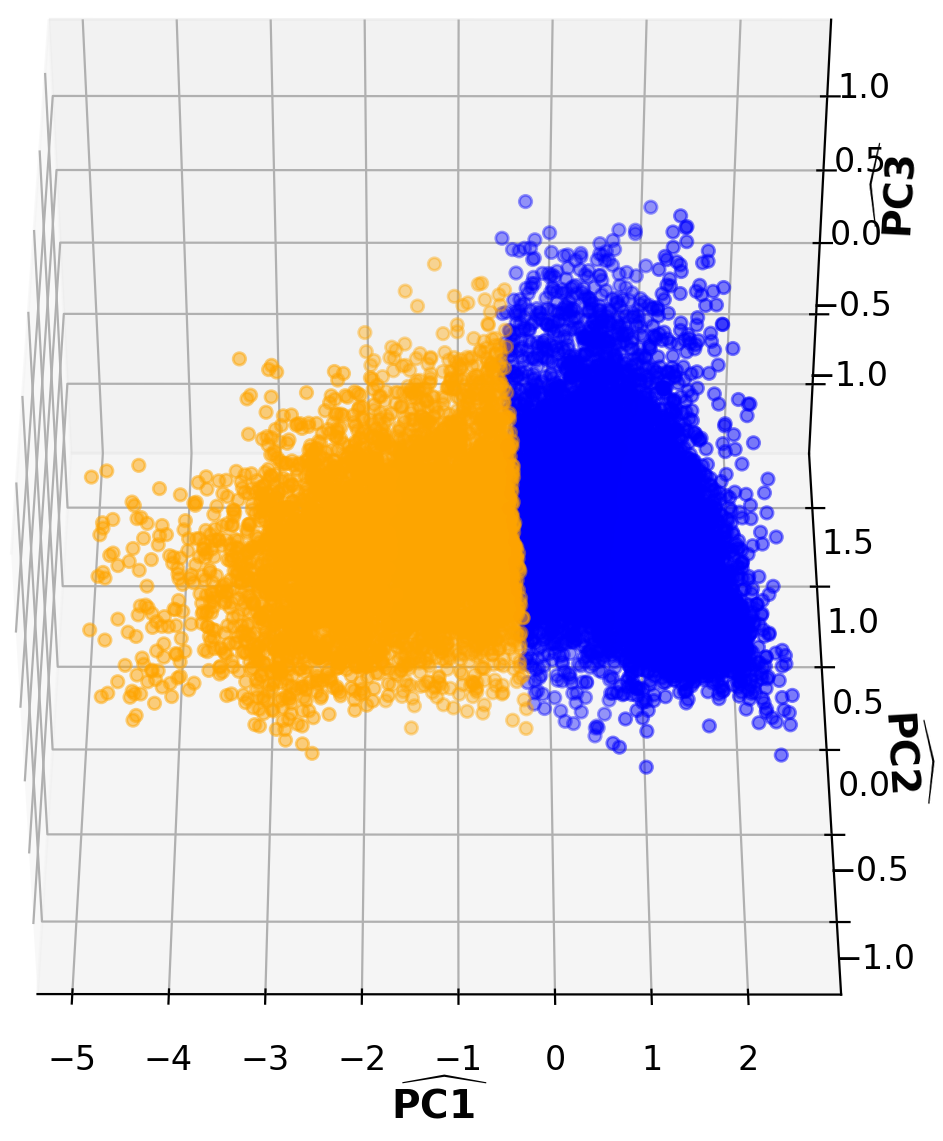}
\caption{direct mapping}
   \label{fig:sm5a}
\end{subfigure}
\hfill
\begin{subfigure}{0.35\textwidth}
\includegraphics[width=1.8in]{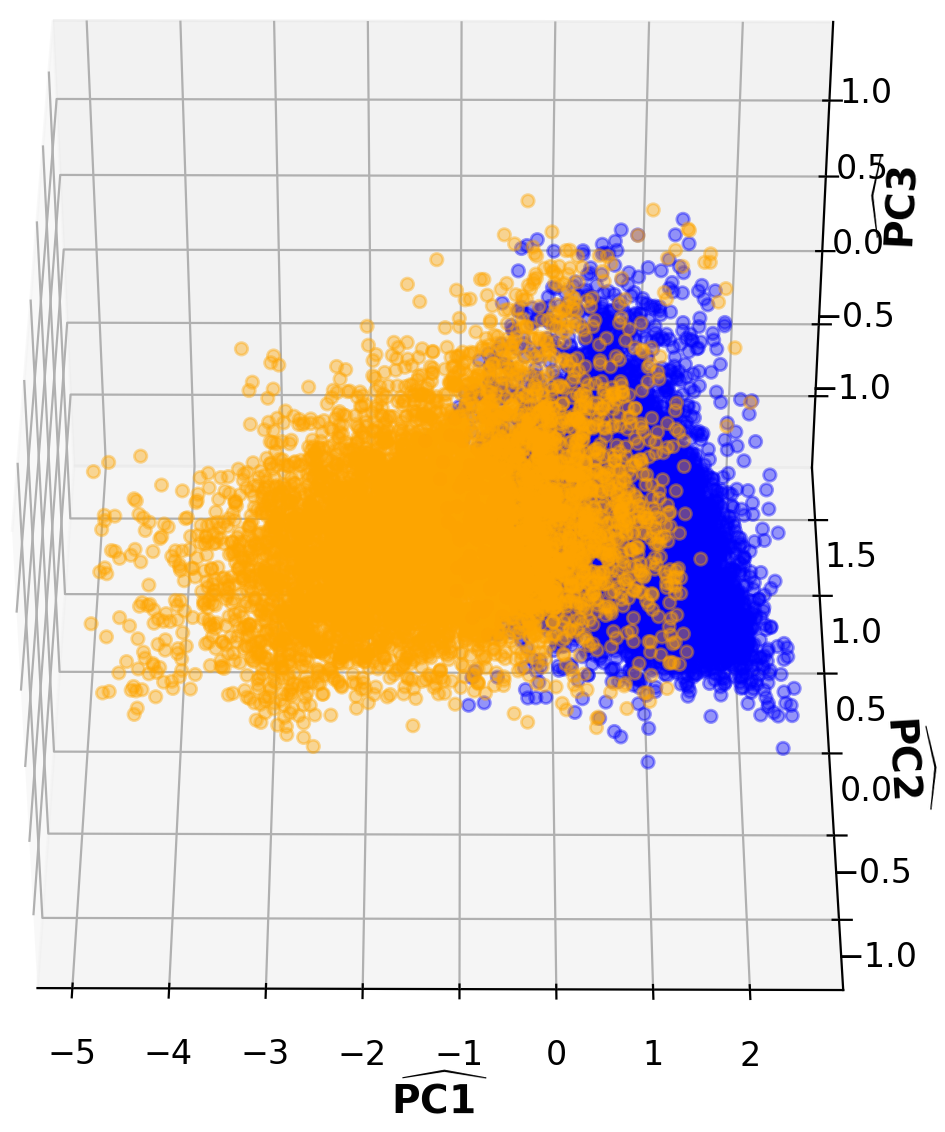}
\caption{converged iteration}
   \label{fig:sm5c}
\end{subfigure}
\hfill
}

\resizebox{\columnwidth}{!}
{
\begin{subfigure}{0.35\textwidth}
\includegraphics[width=1.9in]{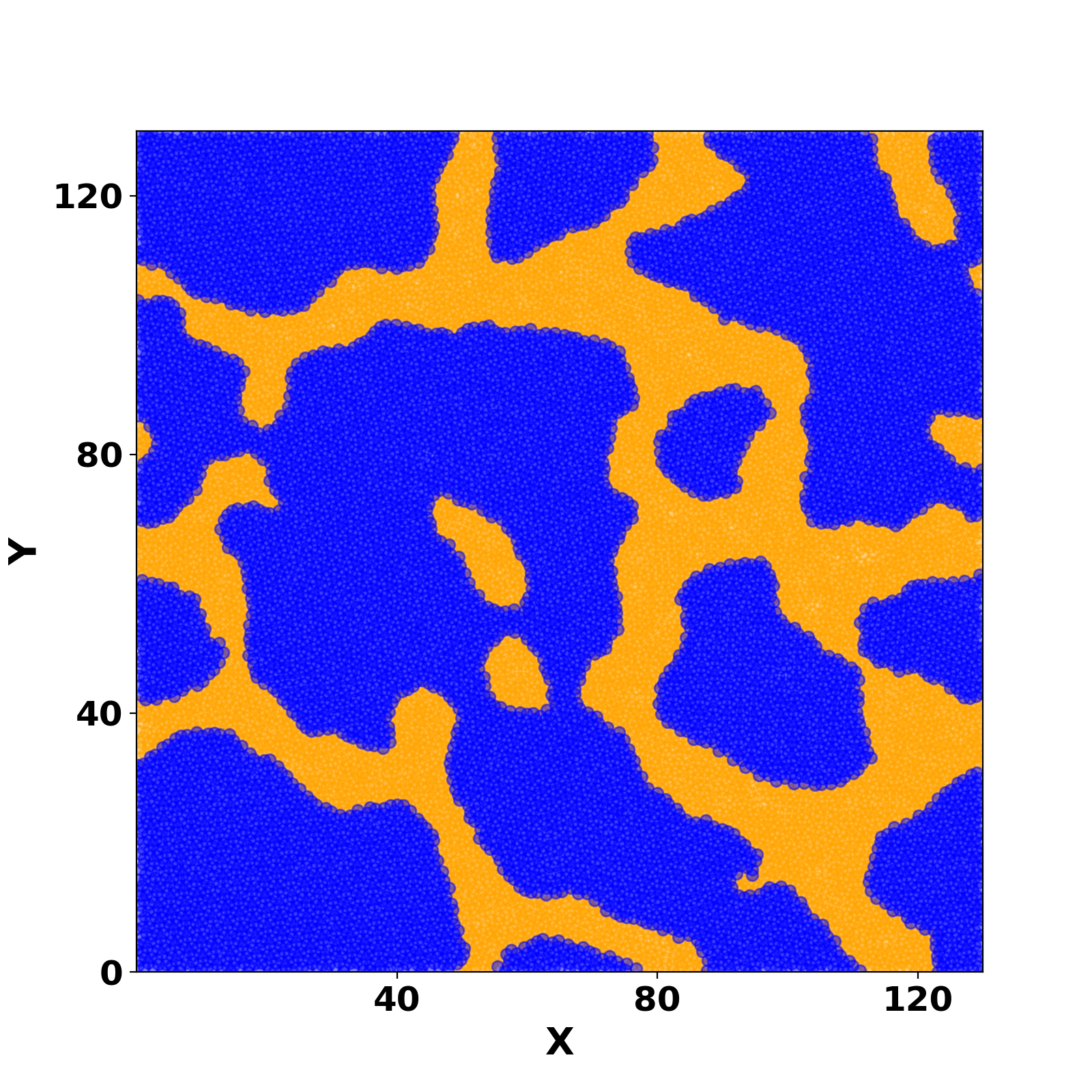}
\caption{direct mapping}
   \label{fig:sm4a}
\end{subfigure}
\hfill
\begin{subfigure}{0.35\textwidth}
\includegraphics[width=1.9in]{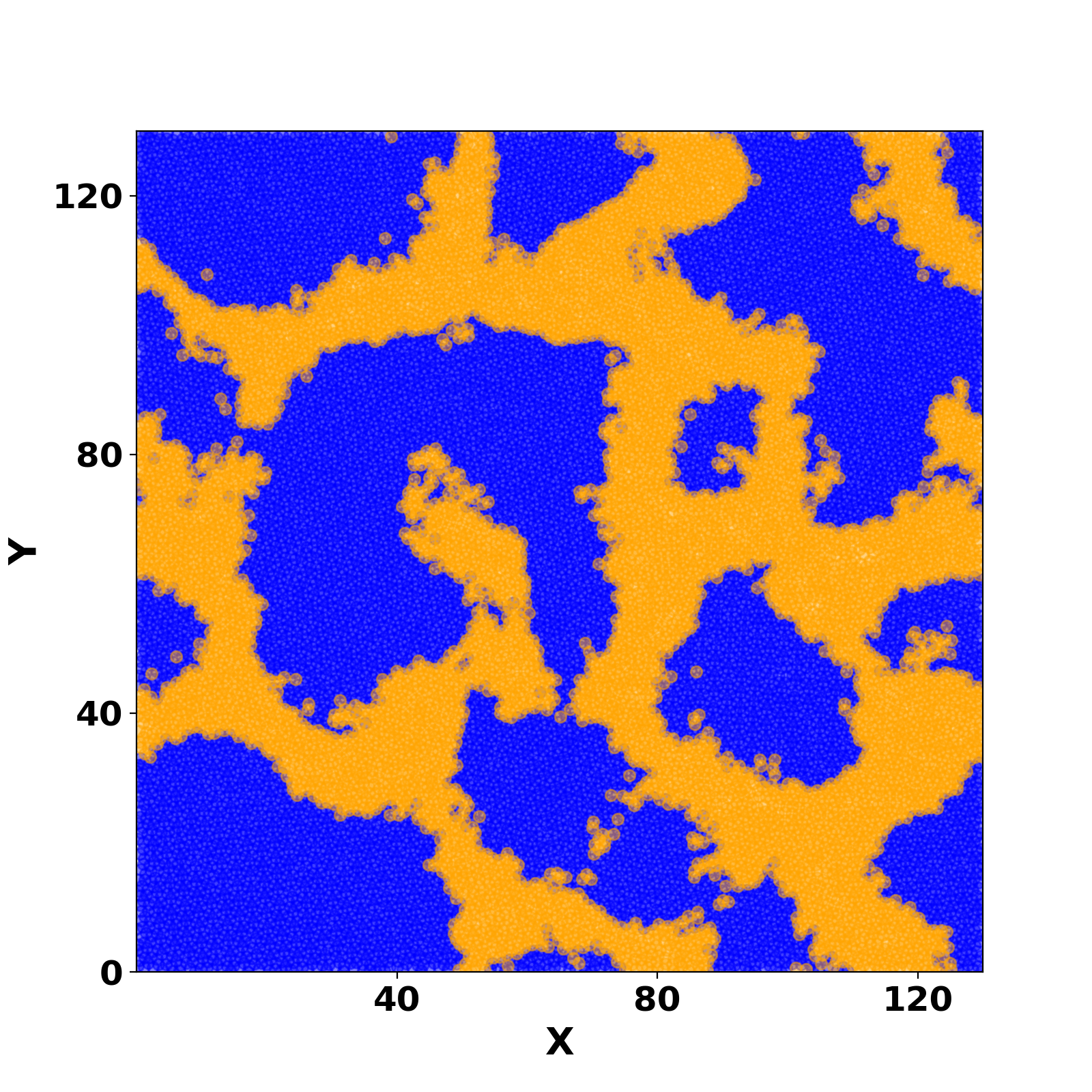}
\caption{converged iteration}
   \label{fig:sm4c}
\end{subfigure}
}

\caption{ A 2D snapshot of 2 meso-states of 16900 particles system at $T^*=0.35$, for both direct mapping and co-learning method. Panels a) and c) are for the PC-space and configurational space from direct mapping while panels b) and d) present clusters formed in the PC-space and configurational space after convergence from co-learning.}  
\label{fig:16900}
%}
\end{figure}

\begin{figure}[h!] 
\resizebox{\columnwidth}{!}
{
\begin{subfigure}{0.35\textwidth}
\includegraphics[width=1.8in]{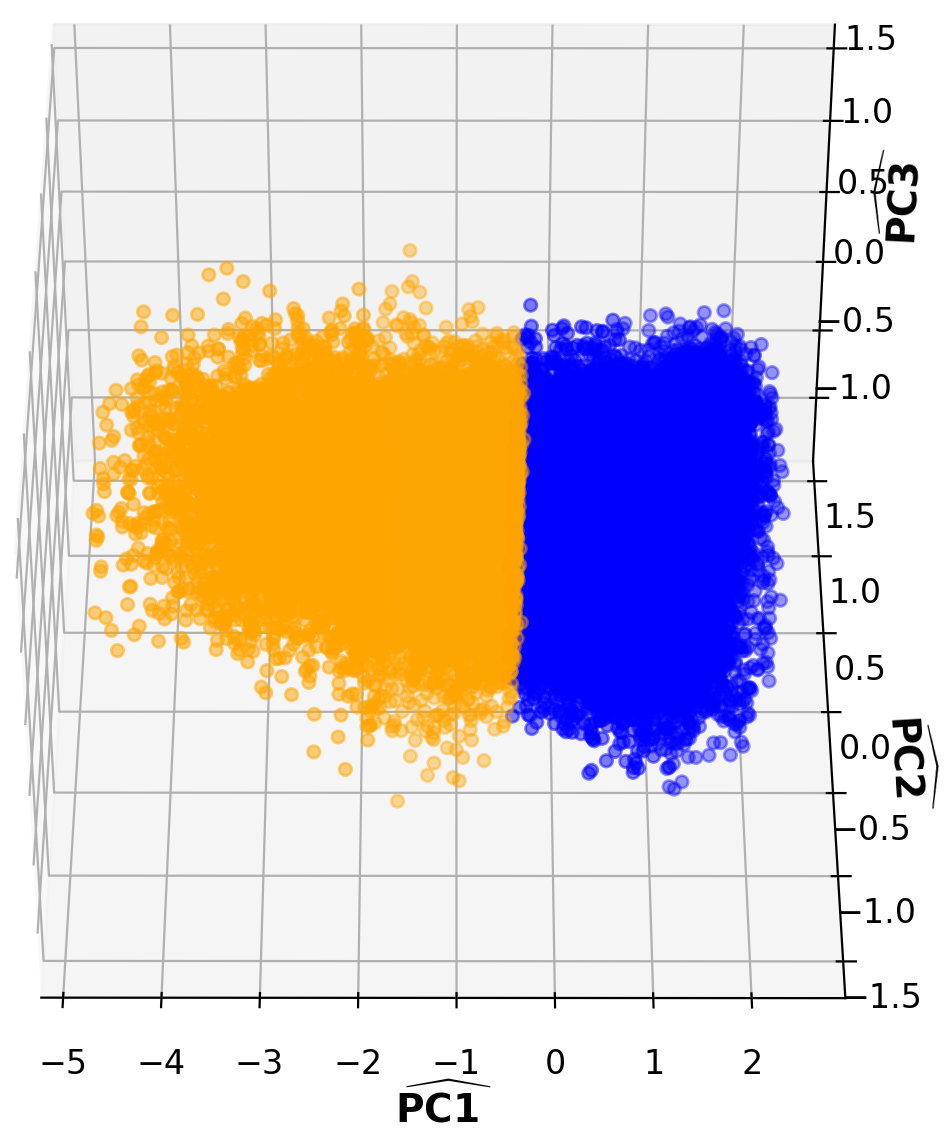}
\caption{direct mapping}
   \label{fig:sm5b}
\end{subfigure}
\hfill
\begin{subfigure}{0.35\textwidth}
\includegraphics[width=1.8in]{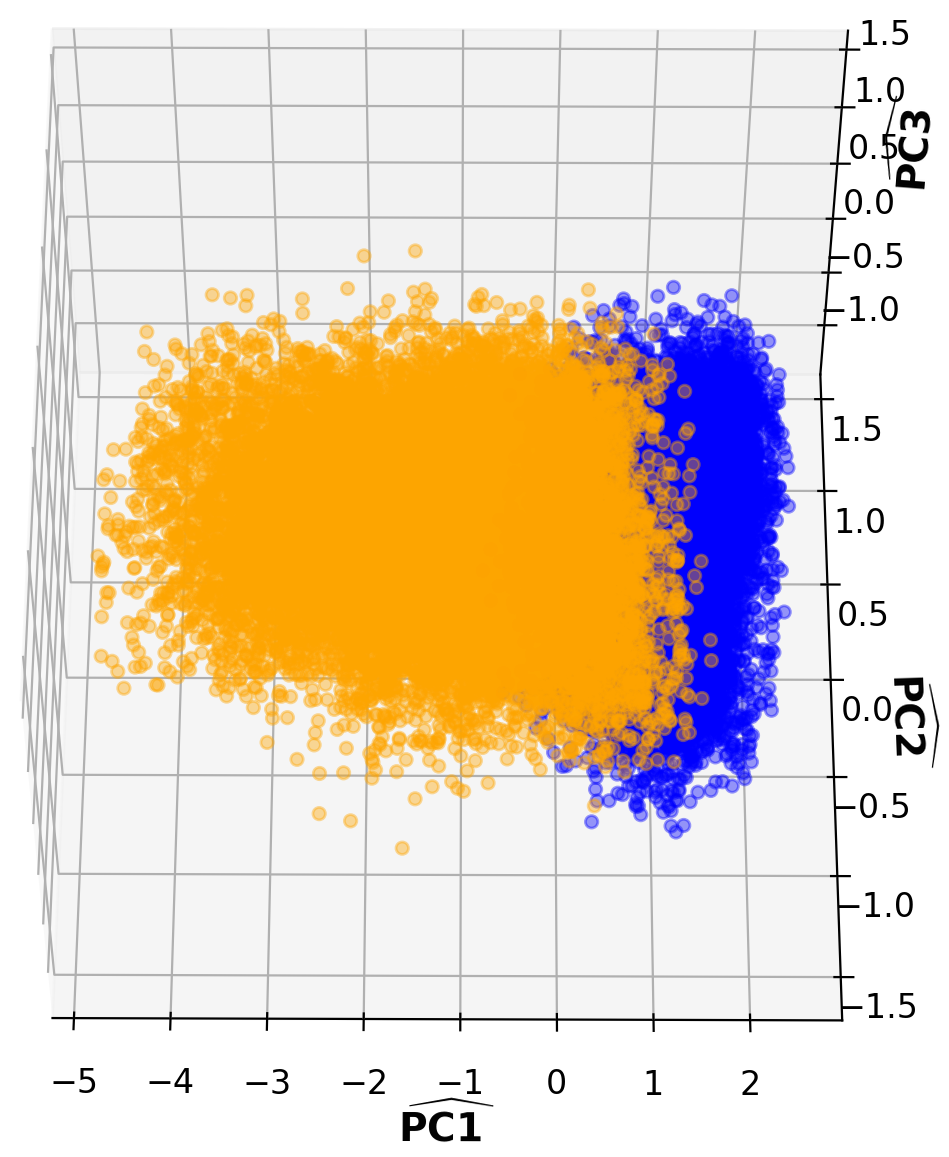}
\caption{converged iteration}
   \label{fig:sm5d}
\end{subfigure}
\hfill
}

\resizebox{\columnwidth}{!}
{
\begin{subfigure}{0.35\textwidth}
\includegraphics[width=1.9in]{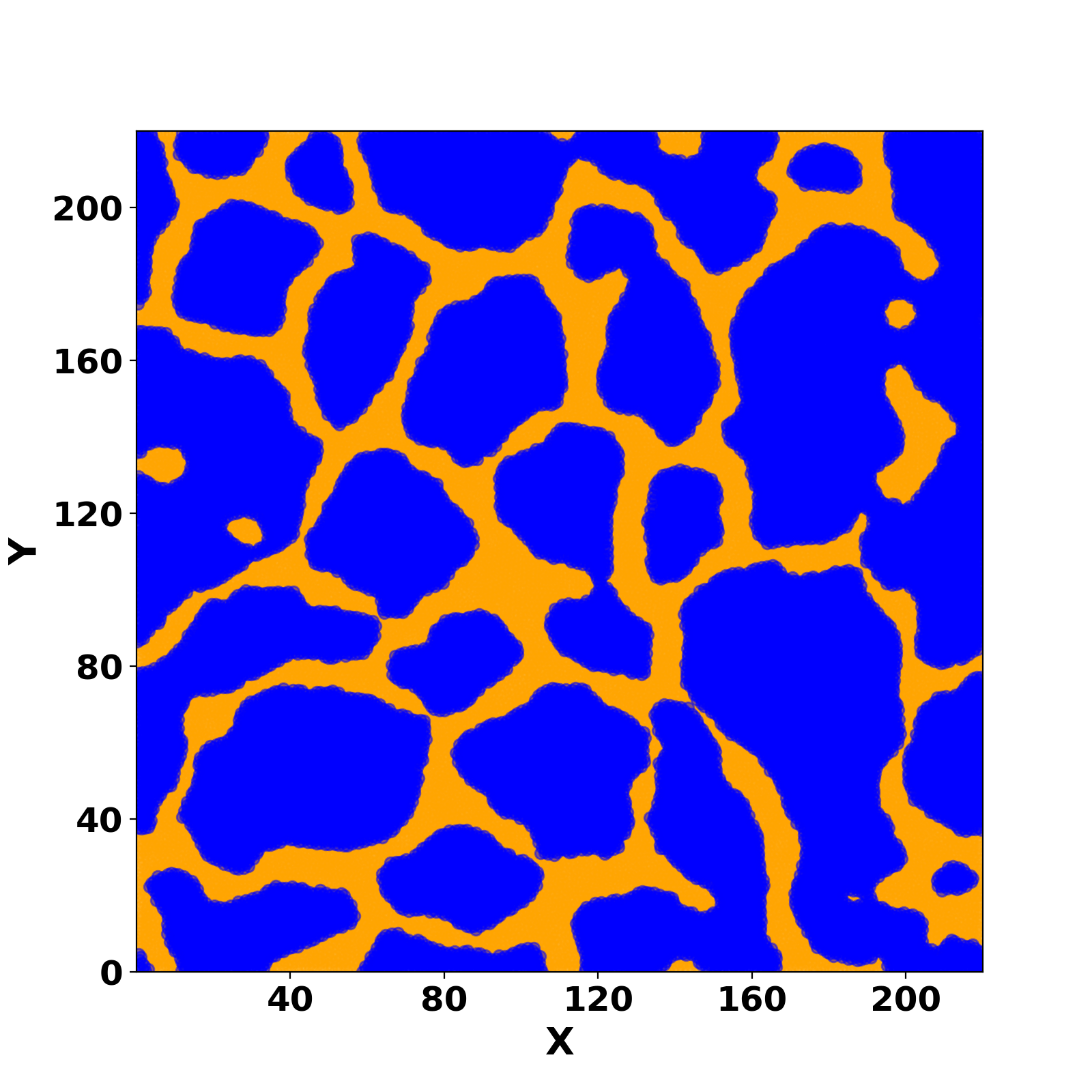}
\caption{direct mapping}
   \label{fig:sm4b}
\end{subfigure}
\hfill
\begin{subfigure}{0.35\textwidth}
\includegraphics[width=1.9in]{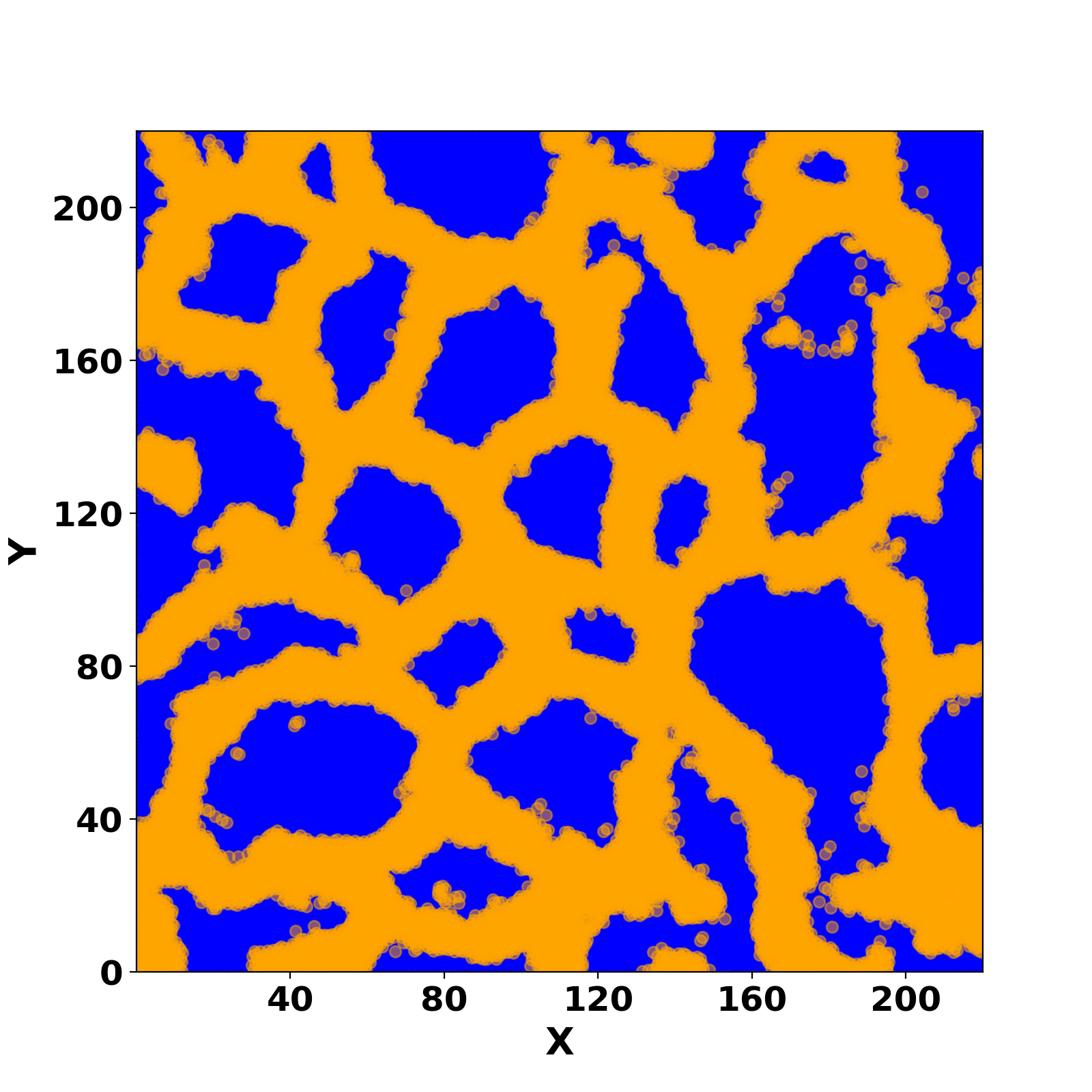}
\caption{converged iteration}
   \label{fig:sm4d}
\end{subfigure}
}

\caption{ A 2D snapshot of 2 meso-states of 48400 particles system at $T^*=0.35$, for both direct mapping and co-learning method. Panels a) and c) are for the PC-space and configurational space from direct mapping while panels b) and d) present clusters formed in the PC-space and configurational space after convergence from co-learning.}  
\label{fig:48400}
%}
\end{figure}

\subsection{Principle Component Analysis and clustering }
Principal Component Analysis (PCA)~\cite{BishopChristopherM2006Pram} is used for dimension reduction, namely to reduce  {\textbf  N}  particles' features to a few correlated ones. The new coordinates (PC's representation) which are generated from the inner product of original $\overline{WCNs}$ matrix with PC's basis (PC-space), are used for K-means clustering to detect meso-states in the structural space. 
PCA is performed by the following steps: 

\begin{itemize}
\item Obtaining the mean-free data $\bold{X = \widetilde{X} - \langle \widetilde{X} \rangle}$ where the average is over ${M}$ particles for each component of WCNs.
\item Forming the correlation matrix $\bold{C = X^\intercal X}$, which is $N\times N$.
\item The principle components $\bold{u_i}$ are obtained after solving the eigenvalue problem: $\bold{Cu_i = {\bm{\sigma_i}}^2u_i }$. The eigenvalue $\bold{\bm{\sigma_i^2}}$ measures the variance of the data along each principle component(PC) $i$. PCA is optimal in term of seeking small numbers of PCs but maximizing cumulative proportion of variance explained (PVE) $\bold{\bm{\sigma_i^2}}$ by each principle component. In other words, the numbers of retained PCs depend on their total PVE such that the total PVE explains $\ge$ 95$\%$ of total variances presented in ${\widetilde{\bf X}}$.
\end{itemize}
The PC's basis is $\bold{U = [u_1,u_2,..u_N]}$ where each $\bold{u_i}$ is a collective coordinate with $N$ components corresponding to the number of features in the data input and the new PC representation mathematically  represented as $\bold{Y= U^\intercal \widetilde{\bf X}}$.

Despite K-means is well-known for unsupervised classification~\cite{F.Noe-pyemma}, it requires a good initial guess of possible numbers of clusters K, hence an implementation of an Elbow convergent test provides a range of number of clusters K effectively where K = 2 to 4 is sufficient enough to serve our purpose. An alternative solution could be an implementation of Bayesian algorithms~\cite{BishopChristopherM2006Pram}. After a careful trial-and-error process, K = 2 is selected. K-means clustering is implemented to classify PC-representation (\ref{fig:a}) into 2 distinct meso-states, then clusters in the configurational space can also be obtained by mapping identities of particles in each meso-state directly shown in \ref{fig:c}. Although a direct mapping of particles associated with a meso-state in the PC-space also results in a domain in real space, the classification of interfacial particles among domains remains uncertain. Initially, minimizing euclidean distance to the centroid is a measure to assign particles to a meso-state in K-means clustering, hence K-means works well to determine membership of particles whose distance are well-separated like core particles of each meso-state. Meanwhile, assignment of interfacial particles could be inclusive due to small differences in the distance of interfacial particels to either states. In other words, the membership of interfacial particles are linearly weighted combination of both states, so a hard assignment using K-means could lead to misclassifications. Therefore, an alternative soft clustering algorithm like Gaussian Mixture is used to improve classification of interfacial particles. To transfer information between the structural space and the configurational space consistently, a co-learning strategy is implemented:

\begin{enumerate}
\item Perform K-means clustering in the PC-space.
\item Use the initial knowledge of the clustering  from the PC-space (direct mapping) to perform Gaussian Mixture (GM) classification to soften the hard assignment of K-means in the configurational space:
\begin{enumerate}
	\item Directly map all particles corresponding to each of PC's states to Cartesian coordinate to identify distinct aggregated domains in the real space.
	\item Build a mixture model of multivariate Gaussian distributions of domains, then maximizing Gaussian probability to assign the membership of each particle to a domain .
\end{enumerate}
\item Repeat the step number 2 but in the PC space: perform GM in the PC-space from the clustering knowledge in the configurational space.
\item Perform GM classification iteratively in both spaces until converged . 
\end{enumerate}

This co-learning strategy is highly non-linear because it gains missed information from a direct transfer between two spaces and consistently re-assigned interfacial particles by utilizing a probabilistic clustering. After a few iterations, all meso-states in both spaces are formed and converged as shown in \ref{fig:b} for PC-space and in \ref{fig:d} for configurational space which present clear evidences of spatial heterogeneity of the system at a supercooled condition. The consistency of the classification protocol is confirmed by increasing the system size: nano-domains structure are quickly converged and formed in both spaces after few iterations in the (\ref{fig:16900} a-d and \ref{fig:48400} a-d.

\section{Results and Discussion} \label{sec:results}

\subsection{Nature of nano-domains: Statics}  \label{sec:nature}

%\begin{figure}
%\parbox{3.3in}
%{
%		\includegraphics[width=2.1in]{fig3.png}
%		\caption{Weighted partial {\it g(r)} decomposition of each meso-state at $T^* =0.35$.} 
%		\label{fig:partial_gr}
%}
%
%\end{figure}

\begin{figure}[h!] 
\resizebox{\columnwidth}{!}
{
\begin{subfigure}{0.35\textwidth}
\includegraphics[width=2in]{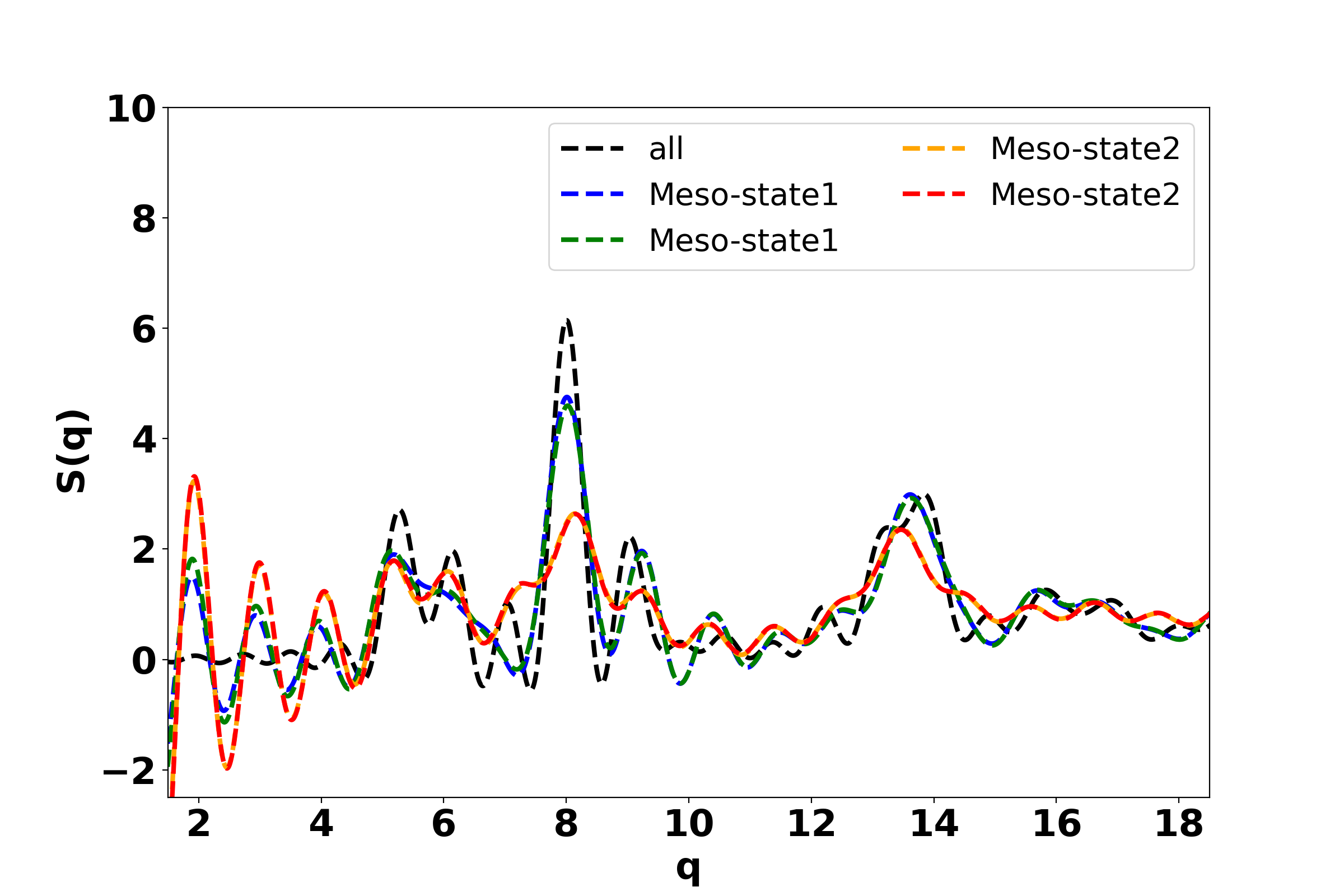}
\caption{partial {\it S(q)}}
   \label{fig:Sq}
\end{subfigure}
\hfill
\begin{subfigure}{0.35\textwidth}
\includegraphics[width=2in]{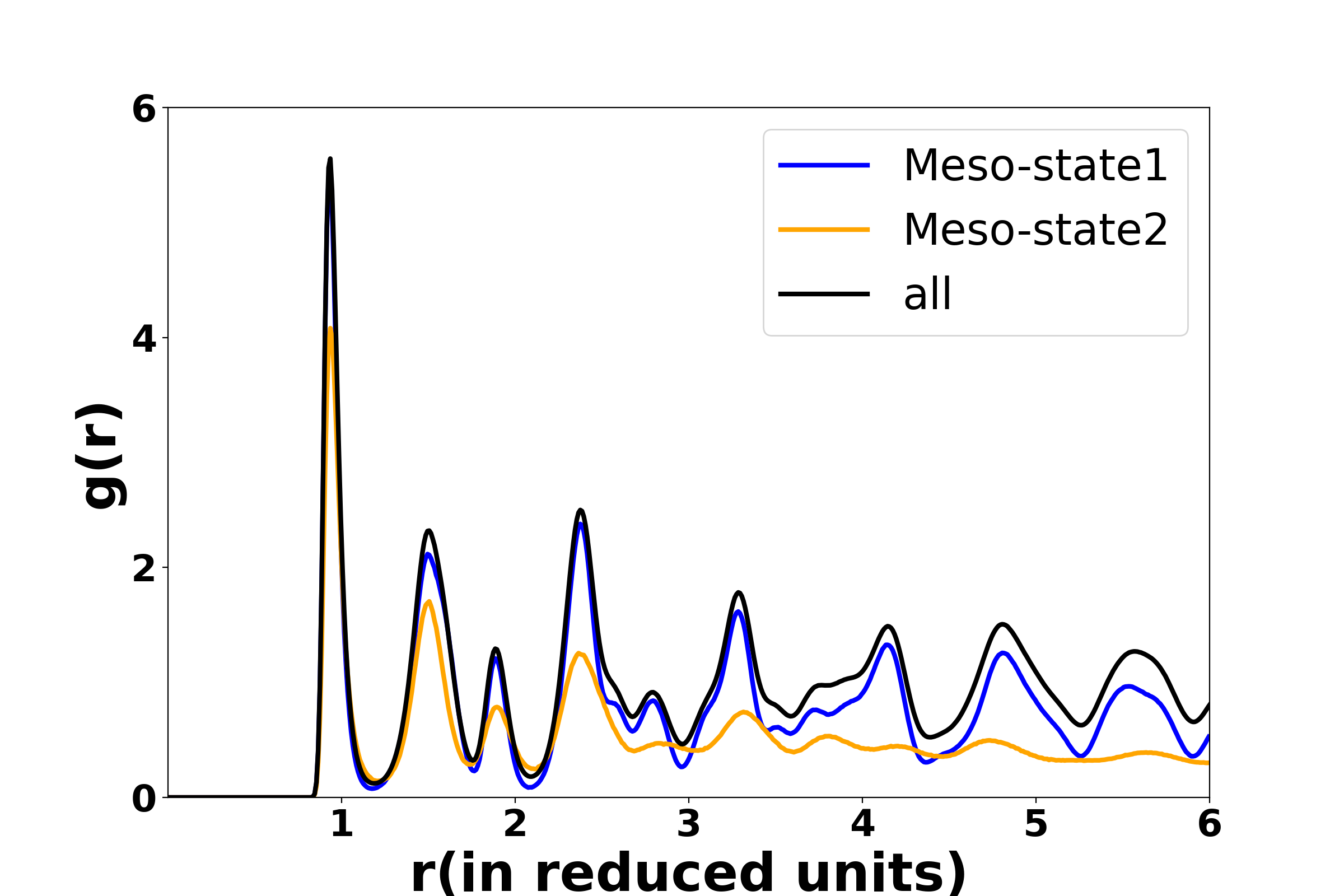}
\caption{partial {\it g(r)} }
   \label{fig:partial_gr}
\end{subfigure}

}
%\parbox{3.3in}
%{
\caption{ (a) Weighted partial static structure factors {\it S(q)}.  (b)  Weighted partial  {\it g(r)}  decomposition of each meso-state at $T^* =0.35$.}  
\label{fig:decomposition}
%}
\end{figure}

To demonstrate non-crystallinity  of meso-states, the static structure factors of two randomly selected domains of the same meso-states for both states and the total system has been calculated. From the observation of all the peaks along the scattering pattern  {\it S(q)} in \ref{fig:Sq}, there is no evidence of crystal structure inside nano-domains. 

\begin{figure}[h!] 
\resizebox{\columnwidth}{!}
{
\begin{subfigure}{0.35\textwidth}
\includegraphics[width=2in]{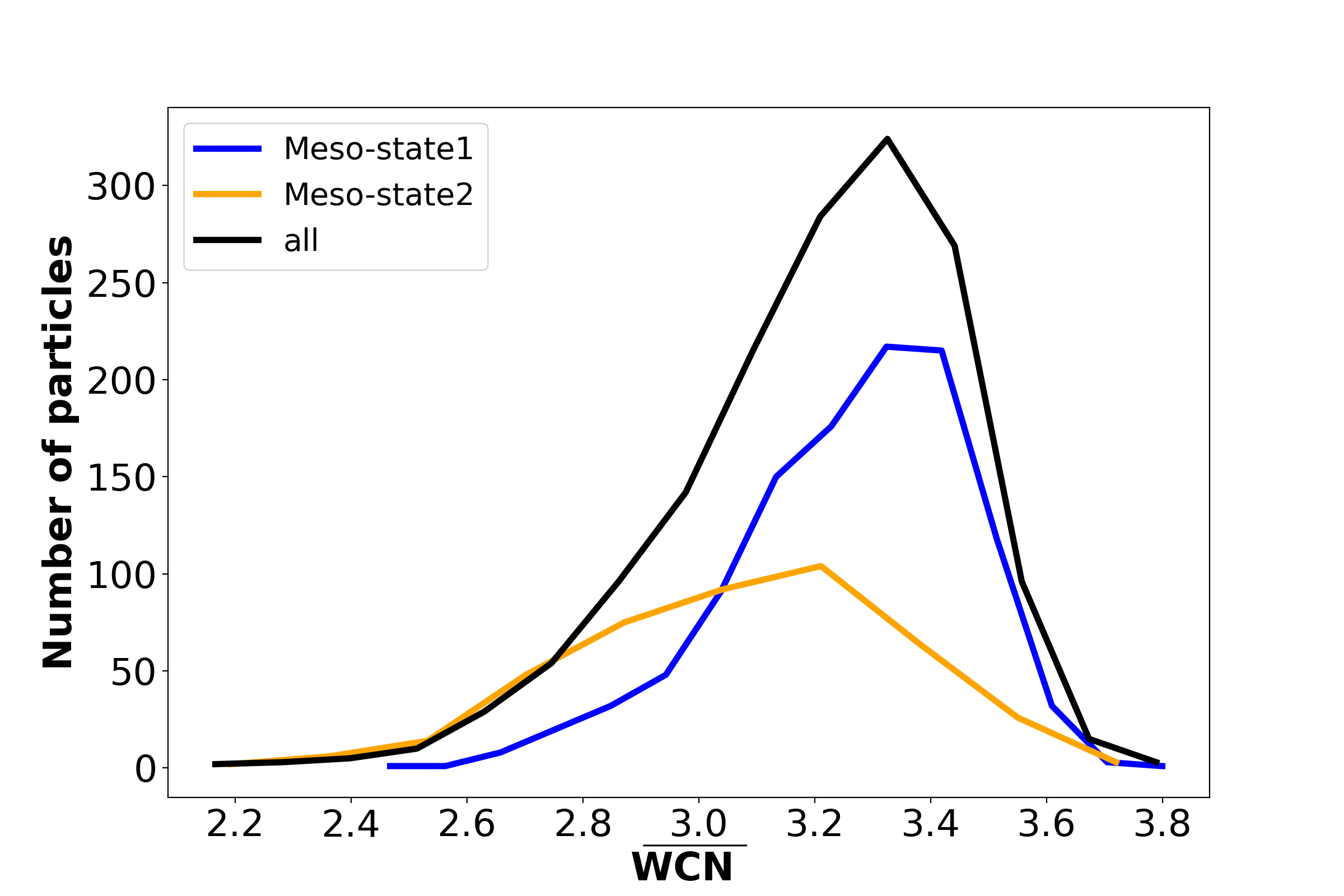}
\caption{ss1}
   \label{fig:1}
\end{subfigure}
\hfill
\begin{subfigure}{0.35\textwidth}
\includegraphics[width=2in]{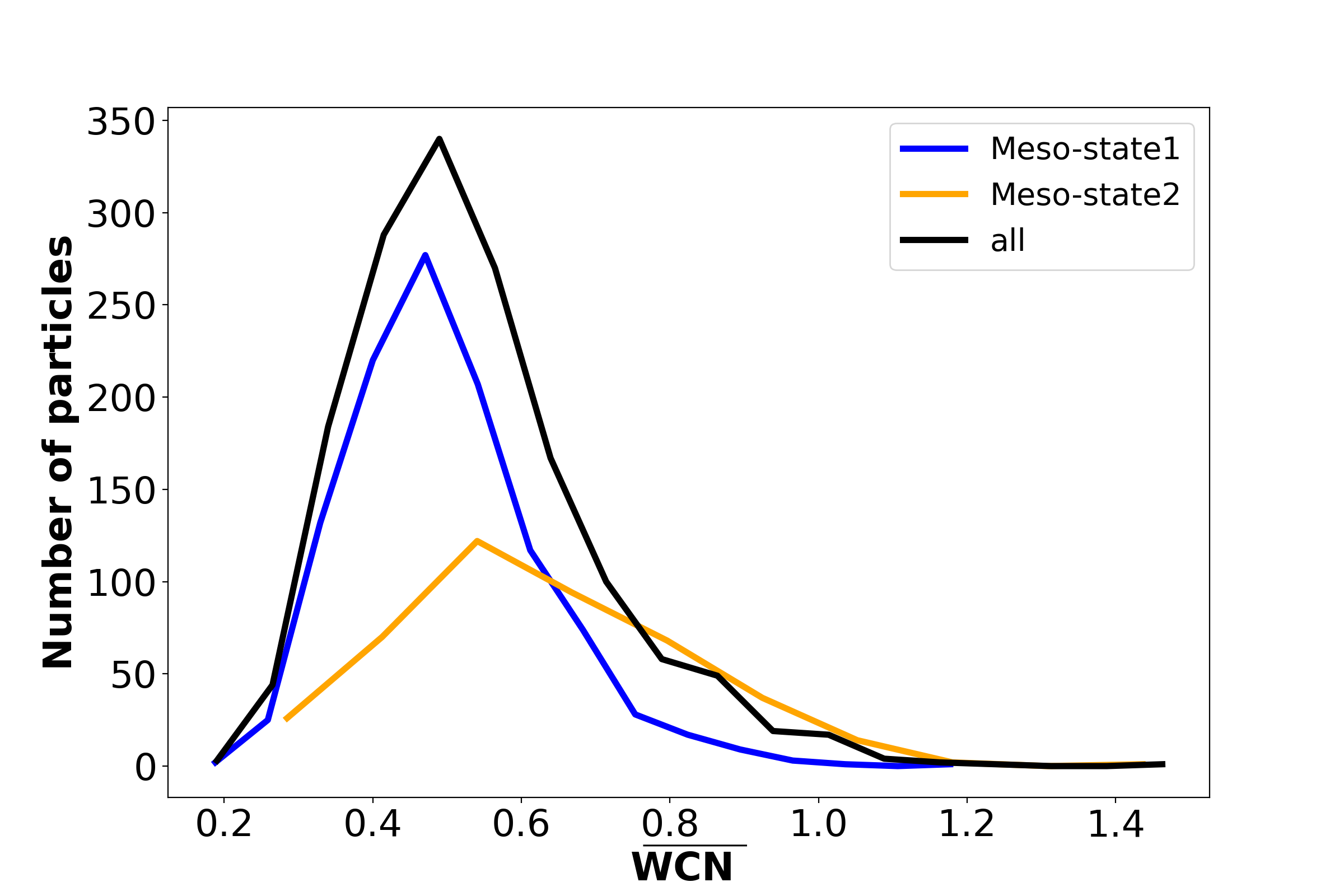}
\caption{ss2}
   \label{fig:2}
\end{subfigure}
\hfill
}

\resizebox{\columnwidth}{!}
{
\begin{subfigure}{0.35\textwidth}
\includegraphics[width=2in]{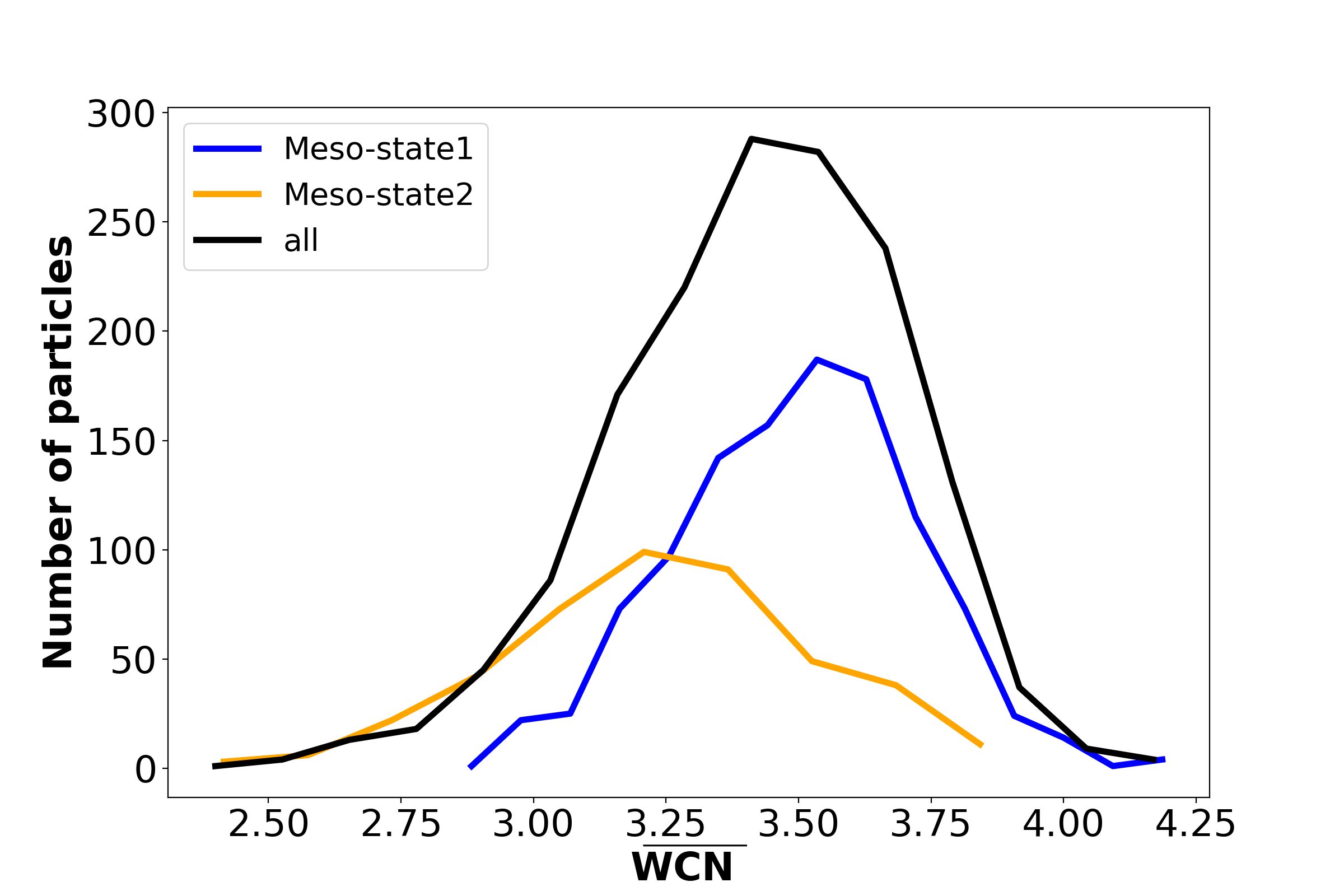}
\caption{ss3}
   \label{fig:3}
\end{subfigure}
\hfill
\begin{subfigure}{0.35\textwidth}
\includegraphics[width=2in]{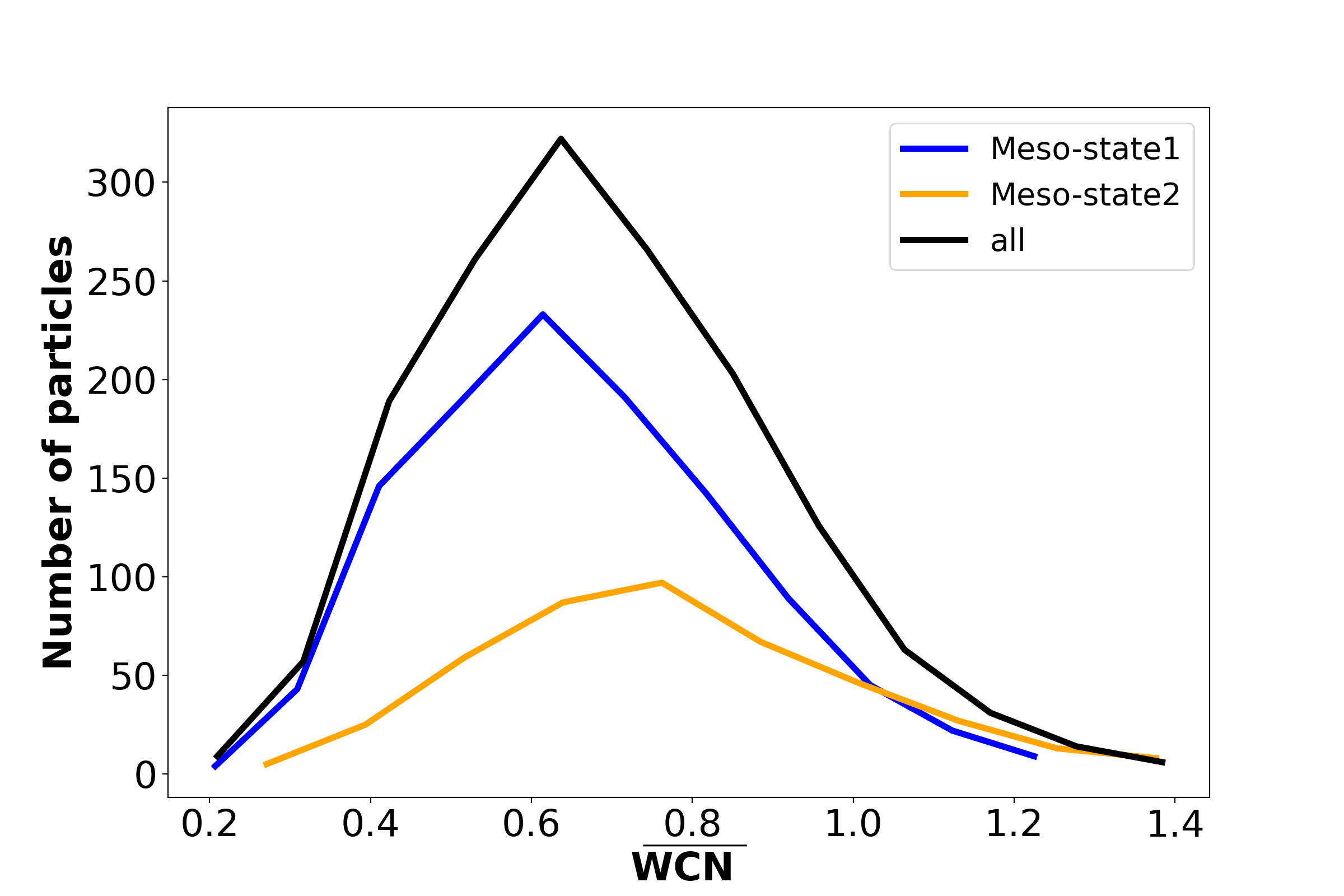}
\caption{ss4}
   \label{fig:4}
\end{subfigure}
}
\caption{ Weighted $\overline{WCNs}$ distributions for the first four solvation shells of each meso-state and whole system at $T^*$ = 0.35, respectively. (ss stands for the solvation shell).}  
\label{fig:wcn_dist}
\end{figure}

The meso-states observed in the PC space show bimodality of $\overline{WCNs}$ distribution along each solvation shell in \ref{fig:wcn_dist}. \ref{fig:wcn_dist} explores the bimodal distribution of $\overline{WCNs}$ along the first four shells; details of more shells are illustrated in \ref{fig:wcn_dist_si} a-f of Appendix \ref{sec:appen}. The presence of a bimodal distribution reflects a coexistence of two meso-states with different local structures. 

From the bimodality of $\overline{WCNs}$, {\it g(r)} for each meso-states are constructed accordingly which is shown in \ref{fig:partial_gr}. The total $\it g(r)$ (black) of the whole system can be decomposed into two partial $\it g(r)$ (blue and orange curves) representing two distinct meso-states smoothed by each solvation shell. The weighted decomposition of $\it g(r)$ is indeed similar to the weighted distribution in $\overline{WCNs}$, thus explains how classification works by classifying  unique local structures of the system.

\begin{figure}
\resizebox{\columnwidth}{!}
{
\begin{subfigure}{0.35\textwidth}
\includegraphics[width=2in]{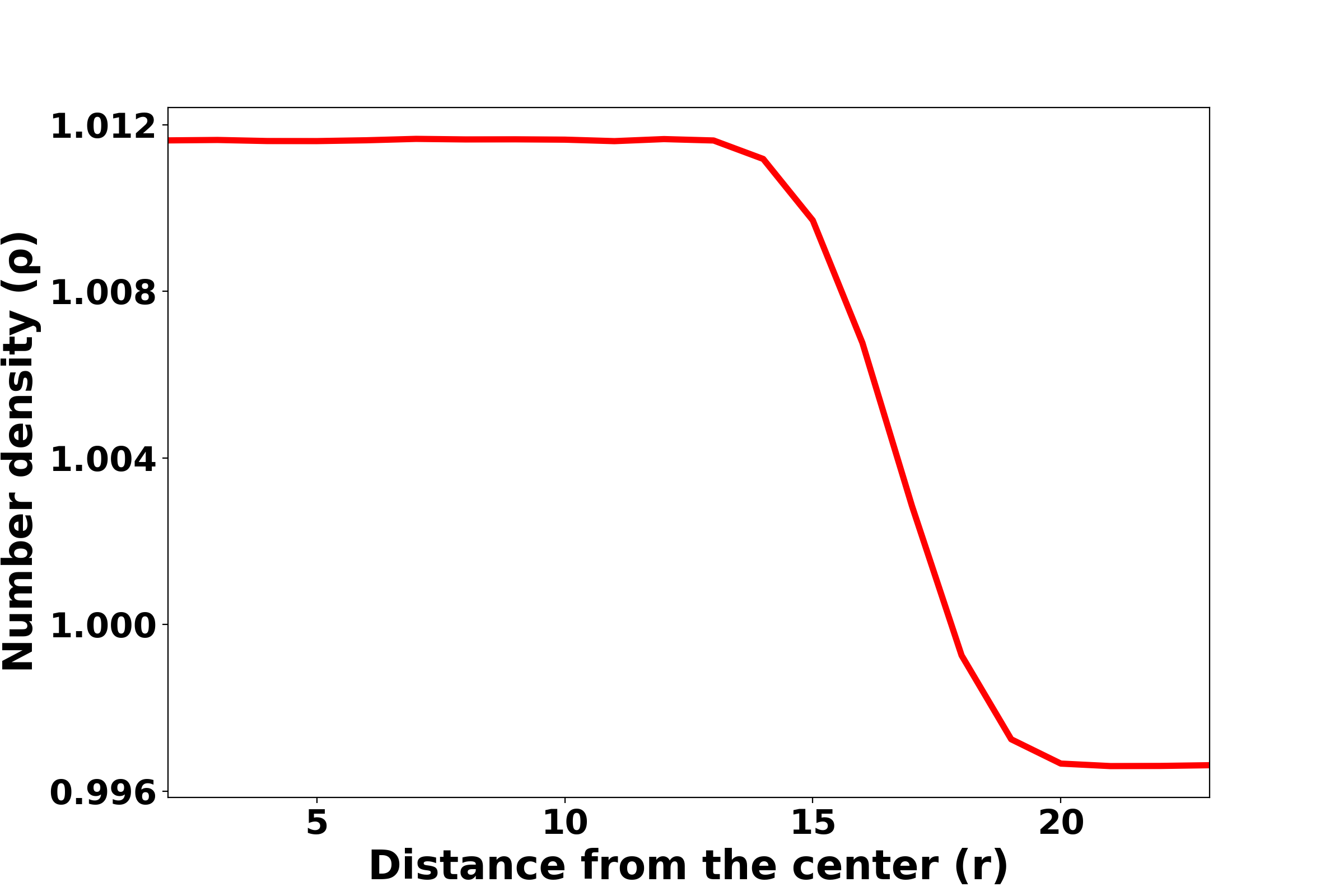}
\caption{Density profile}
   \label{fig:density}
\end{subfigure}
\hfill
\begin{subfigure}{0.35\textwidth}
\includegraphics[width=2in]{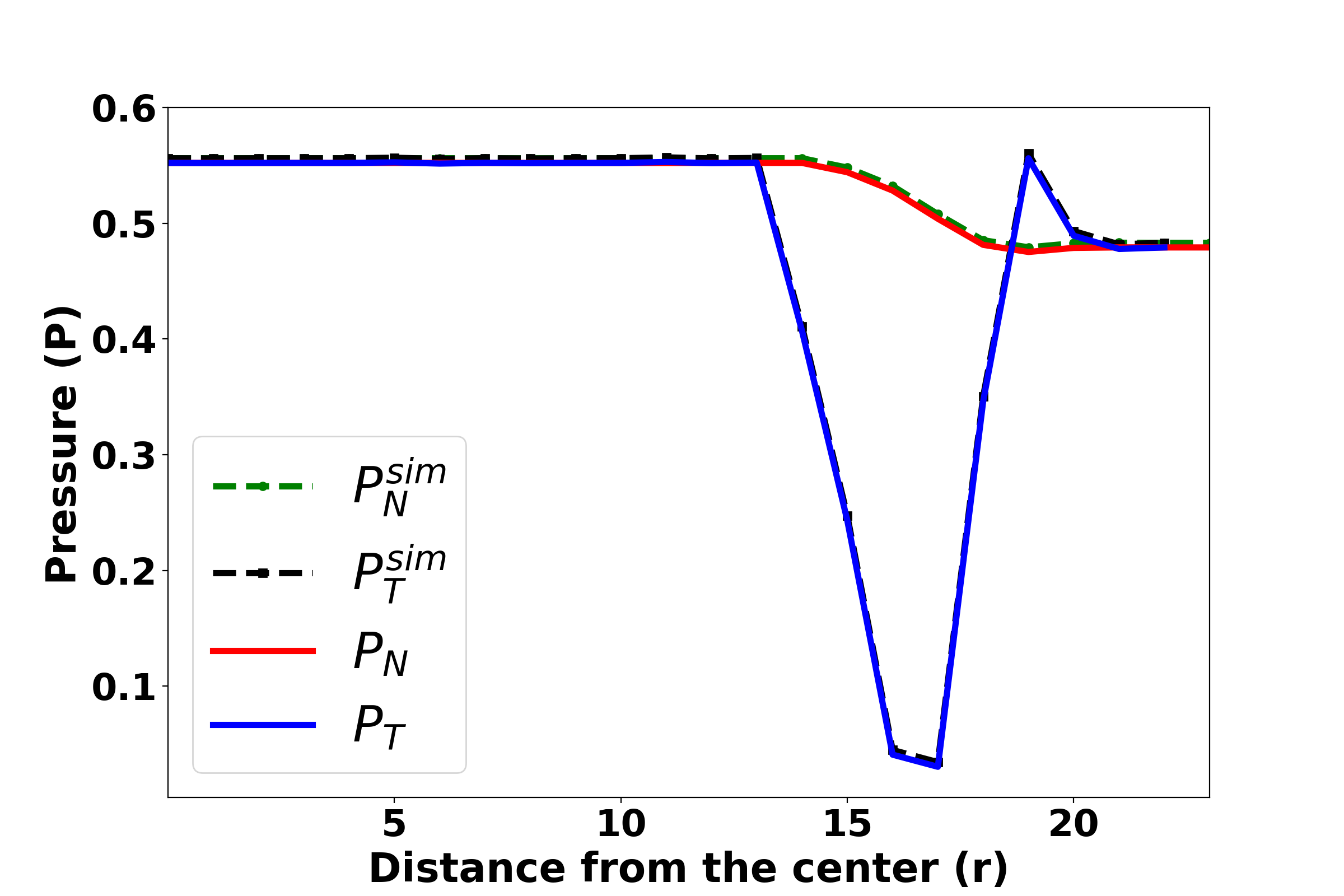}
\caption{Pressure profile}
   \label{fig:press}
\end{subfigure}
}
%\parbox{3.3in}
%{
\caption{ (a) Density profile varying with distance from the center of a patch of meso-state1  at $T^*=0.35$.  (b) Normal and tangential components of the pressure tensor varying with the distance from the center of a patch of meso-state1. The green line and black line are normal ($P_N^{sim}$) and tangential ($P_T^{sim}$) pressures obtained in the simulation, respectively. The red line is the fit to the normal pressure ($P_N$) and the blue line is the tangential pressure ($P_T$) obtained to verify mechanical equilibrium.}  
\label{fig:density_press}
%}
\end{figure}

Another way to quantitatively interpret these distinguished meso-states is to compute their density and pressure profile in the configurational space. Because the domain shapes of meso-state1 are irregular, the thermodynamic properties of the two meso-states such as density and pressure are calculated using a circular region inside a domain of meso-state1 and a shell in the outermost meso-state 2 region as shown in \ref{fig:density_press} along one direction, while other directions give similar results. \ref{fig:density} shows the radial atomic density distribution from the center of a domain of meso-state 1. Meanwhile, the pressure distribution of the system is calculated as described below: The four components of the pressure tensor ($\it p_{xx}$,  $\it p_{yy}$, $\it p_{xy}$ and $\it p_{yx}$) for each atom are computed in the Cartesian coordinate. The pressure tensor is then transformed into polar coordinates whose corresponding components will be ($\it p_{rr}$,  $\it p_{\theta\theta}$, $\it p_{r\theta}$ and $\it p_{\theta r}$). It is noted that the order of magnitude of the off-diagonal terms is the same as the diagonal terms. By default, LAMMPS treats 2D particles as finite-size 3D spheres, hence actually simulated a pseudo-2D system whose 3D-sphered particles are enforced to only move in the $\it x$ and $\it y$ direction while the $\it z$-components of velocities and forces are zeroed out at every time step. This pseudo-2D model physically explains why the off-diagonal terms have the same magnitude order comparing to the diagonal ones. In an exact-2D system, the pressure tensor should have negligible contribution of off-diagonal terms:
\begin{equation} \label{eq:press_comp}
	P(r) = P_N (r)\textbf e_r\textbf e_r + P_T (r)\textbf e_{\theta}\textbf e_{\theta}
\end{equation}
where $\textbf e_r$, $\textbf e_{\theta}$ are unit vectors and $\it P_N$ and $\it P_T$ are the radial or normal and transverse components of the pressure tensor, respectively. The radial profiles of the components $\it P_N(r)$ and $\it P_T(r)$ are obtained by integrating out the angular degrees of freedom over thin cicular shells extending outwards from the origin. Fig.~\ref{fig:press} shows the normal ($\it P_N$) and tangential ($\it P_T$) pressure profiles for the circular interface. It is verified that the normal and tangential profiles statisfy the mechanical equilibrium,$\bm{\nabla } \cdot \bm{P} = 0$, which in polar coordinates is given by: 
\begin{equation} \label{eq:press_mech_eq}
	P_T (r) = P_N (r) + r \frac {dP_N (r)} {dr},
\end{equation}
where the second term is the derivative of the normal pressure with respect to distance from the center of the patch of meso-state1. 

\ref{fig:density_press} indicates the formation of interfaces which signifies the co-existence of two local distinct states. It will be interesting to further characterize such co-existence by computing the chemical potential in both meso-states.

\subsection{ Nature of the nano-domains: Dynamics}

\begin{figure}
\resizebox{\columnwidth}{!}
{
\begin{subfigure}{0.35\textwidth}
\includegraphics[width=2.3in]{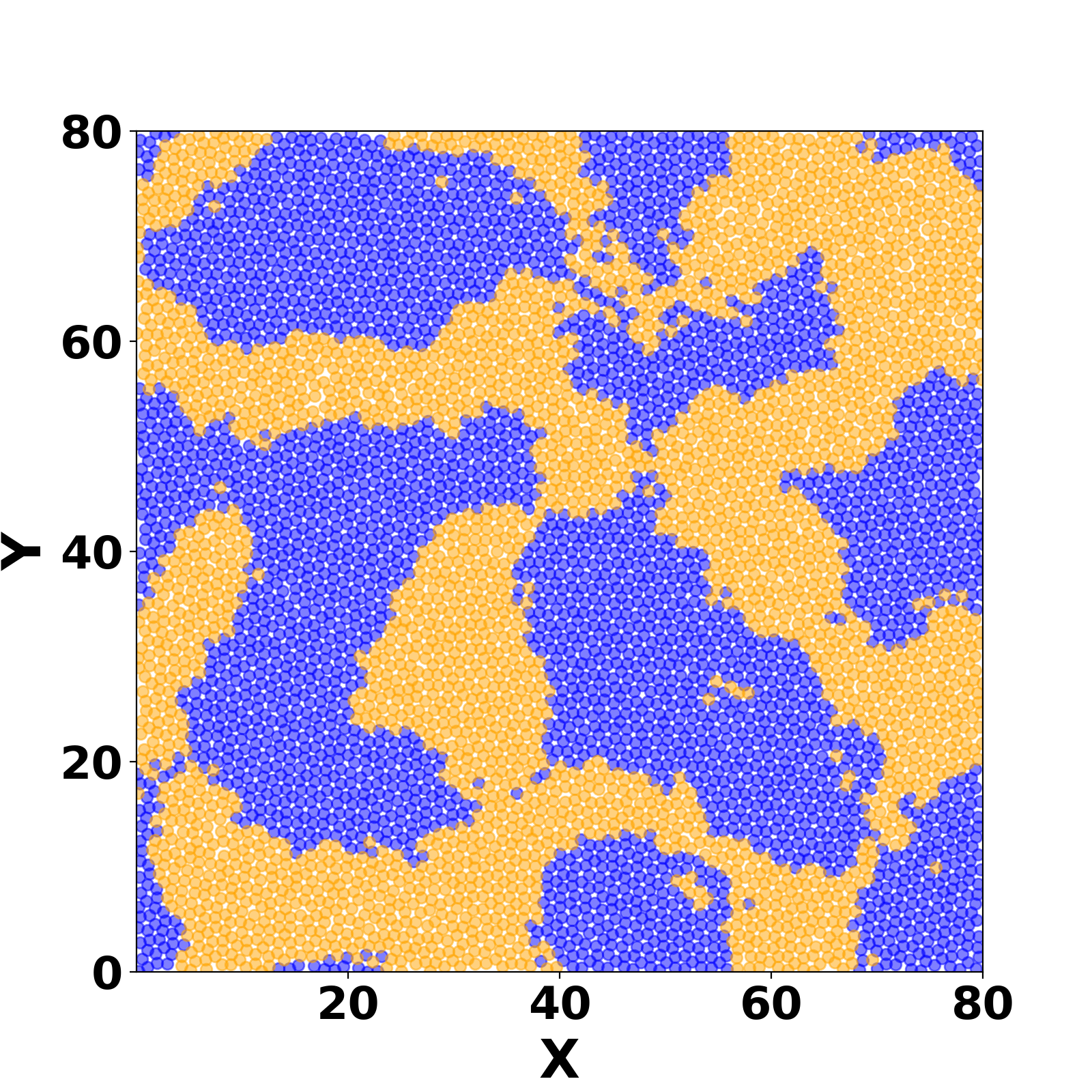}
\caption{$T^*$ = 0.35, Snapshot 1}
   \label{fig:T0.35_0ns}
\end{subfigure}
\hfill
\begin{subfigure}{0.35\textwidth}
\includegraphics[width=2.3in]{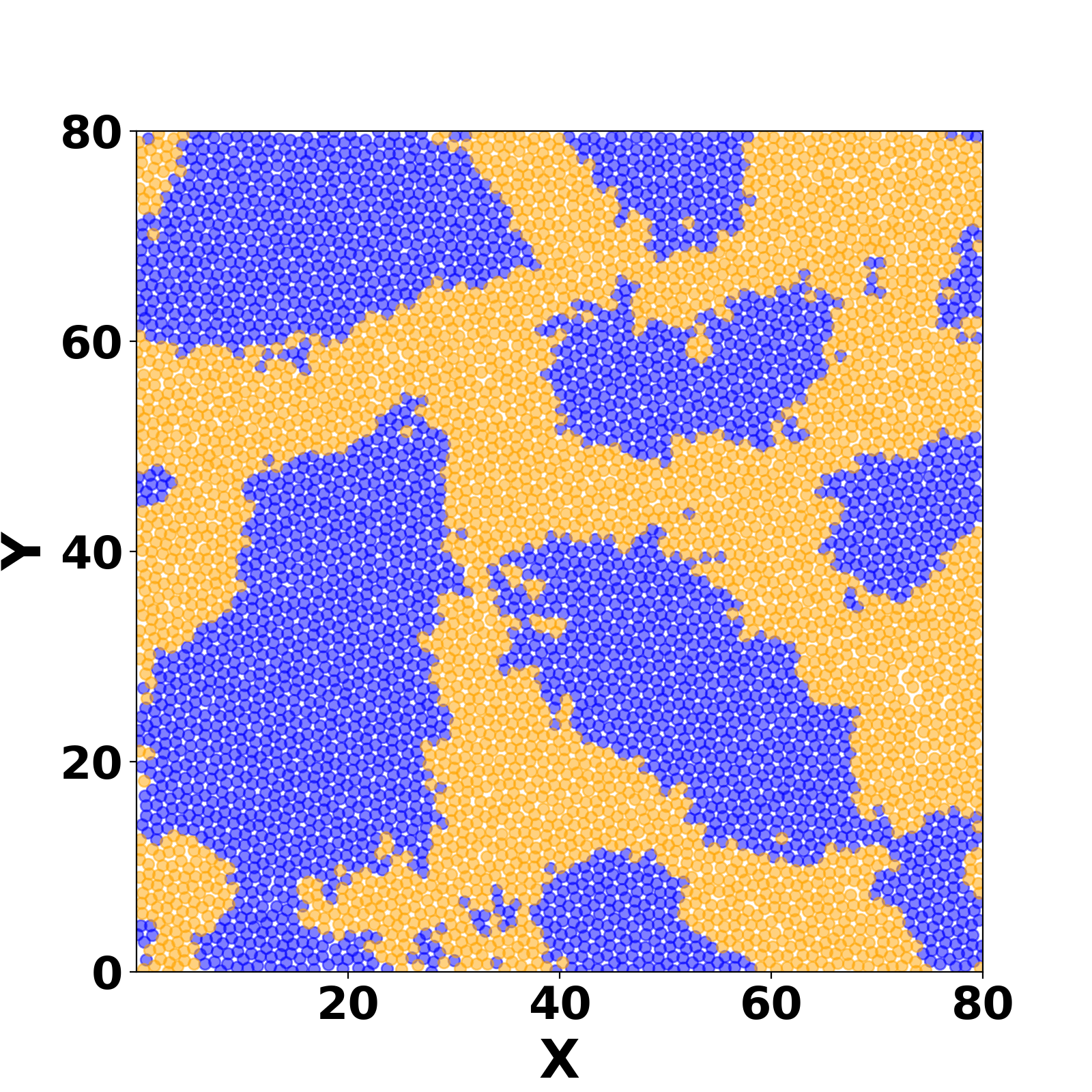}
\caption{$T^*$ = 0.35, Snapshot 2}
   \label{fig:T0.35_10ns}
\end{subfigure}
\hfill
\begin{subfigure}{0.35\textwidth}
\includegraphics[width=2.3in]{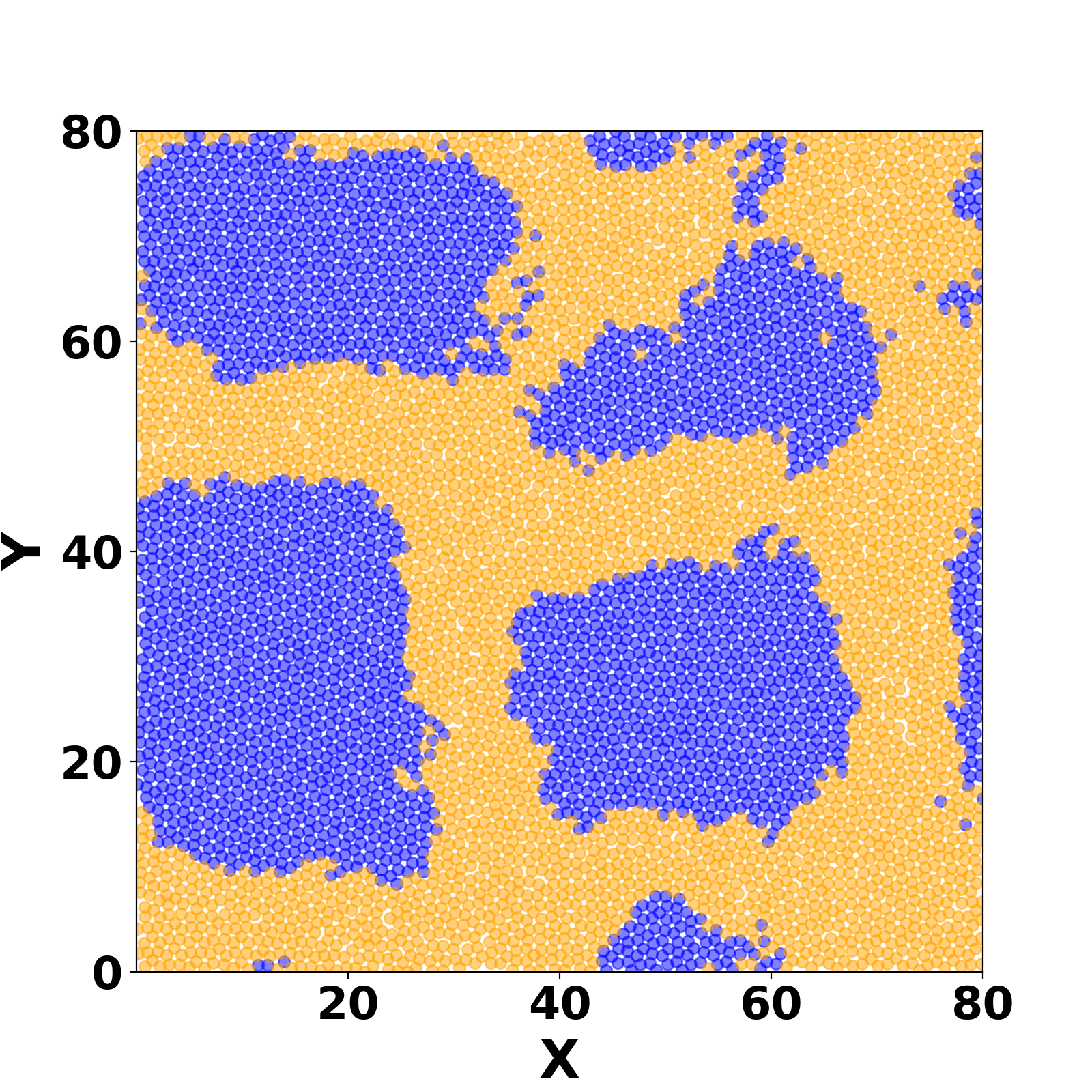}
\caption{$T^*$ = 0.35, Snapshot 3}
   \label{fig:T0.35_20ns}
\end{subfigure}
\hfill
}
\resizebox{\columnwidth}{!}
{
\begin{subfigure}{0.35\textwidth}
\includegraphics[width=2.3in]{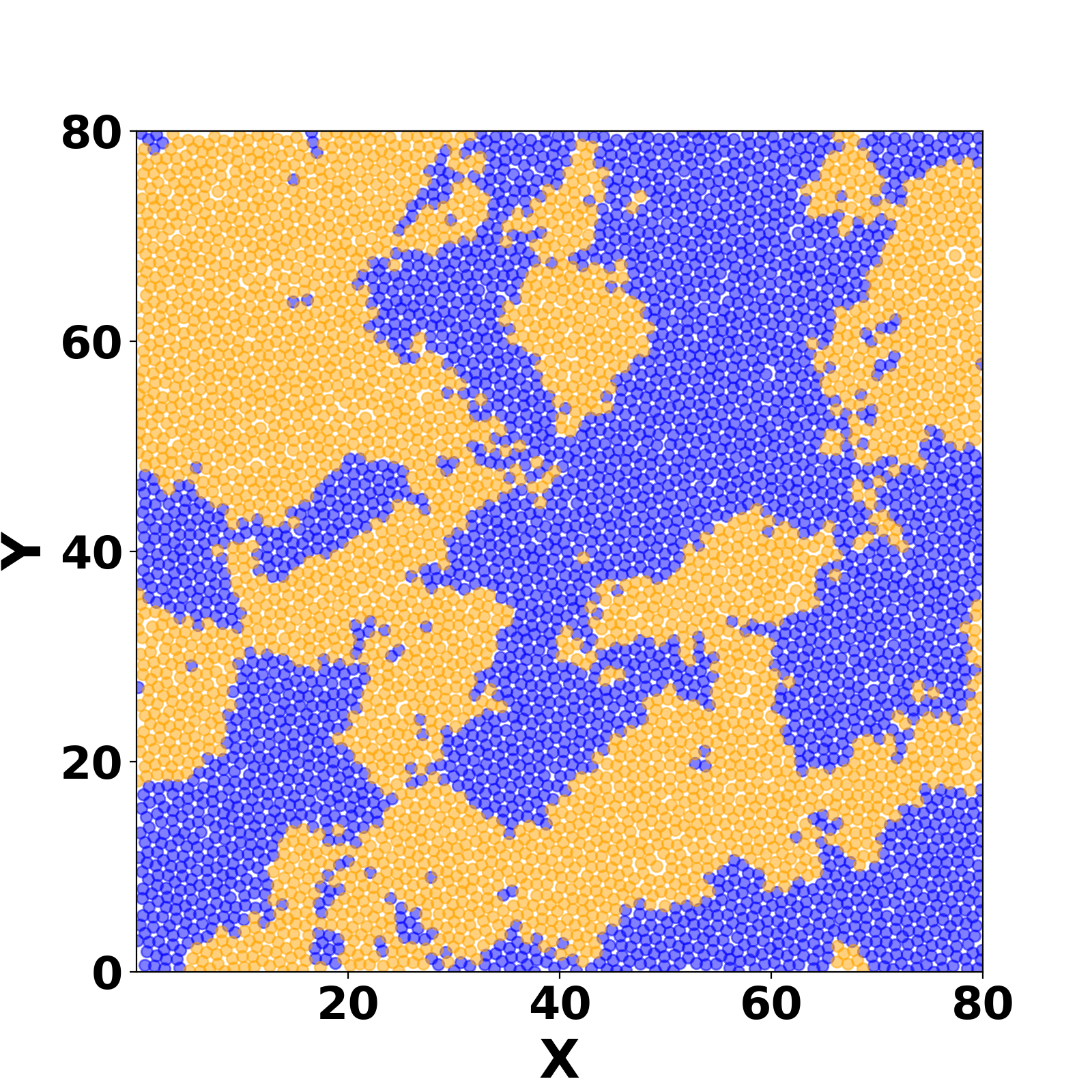}
\caption{$T^*$ = 0.2, Snapshot 1}
   \label{fig:T0.2_0ns}
\end{subfigure}
\hfill
\begin{subfigure}{0.35\textwidth}
\includegraphics[width=2.3in]{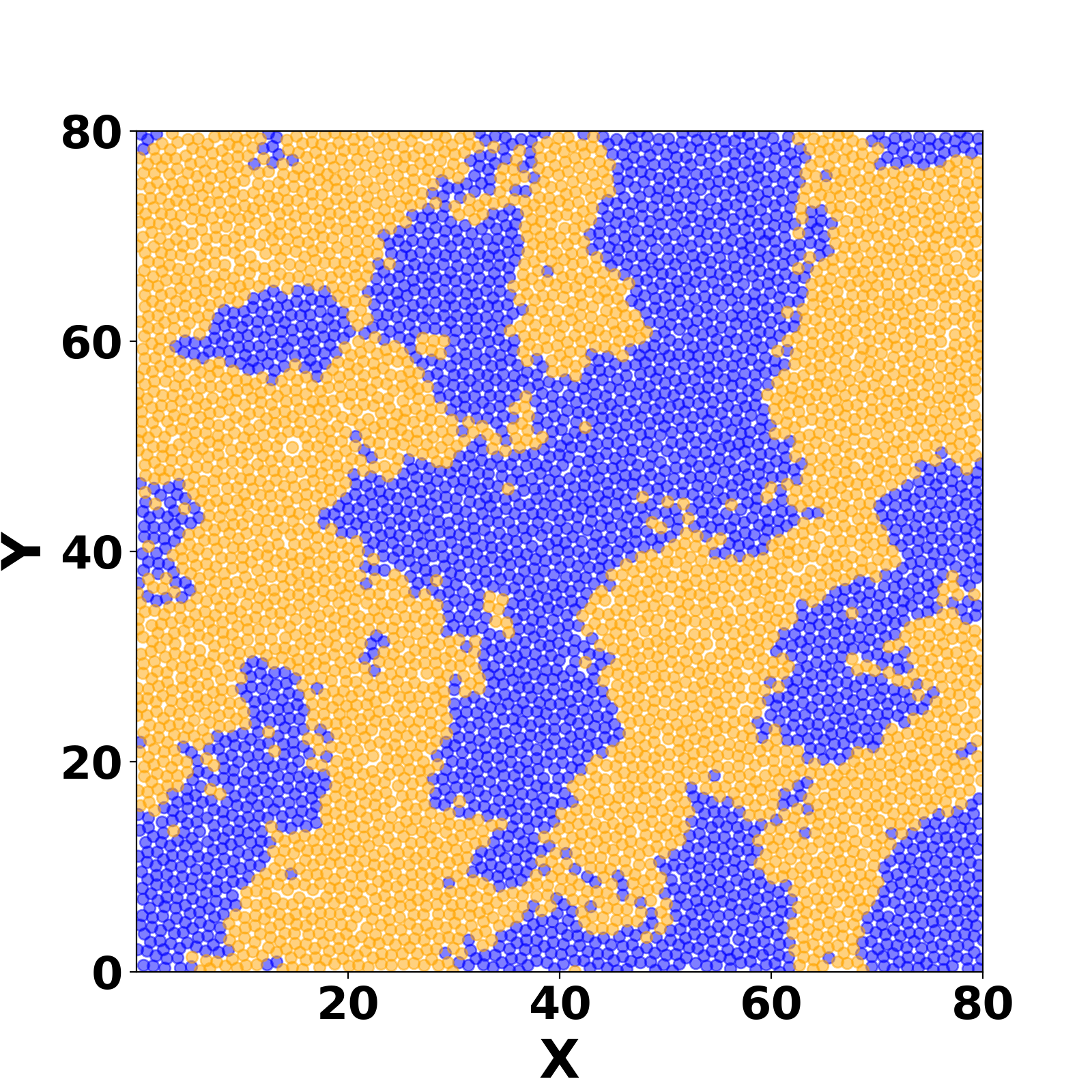}
\caption{$T^*$ = 0.2, Snapshot 2}
   \label{fig:T0.2_10ns}
\end{subfigure}
\hfill
\begin{subfigure}{0.35\textwidth}
\includegraphics[width=2.3in]{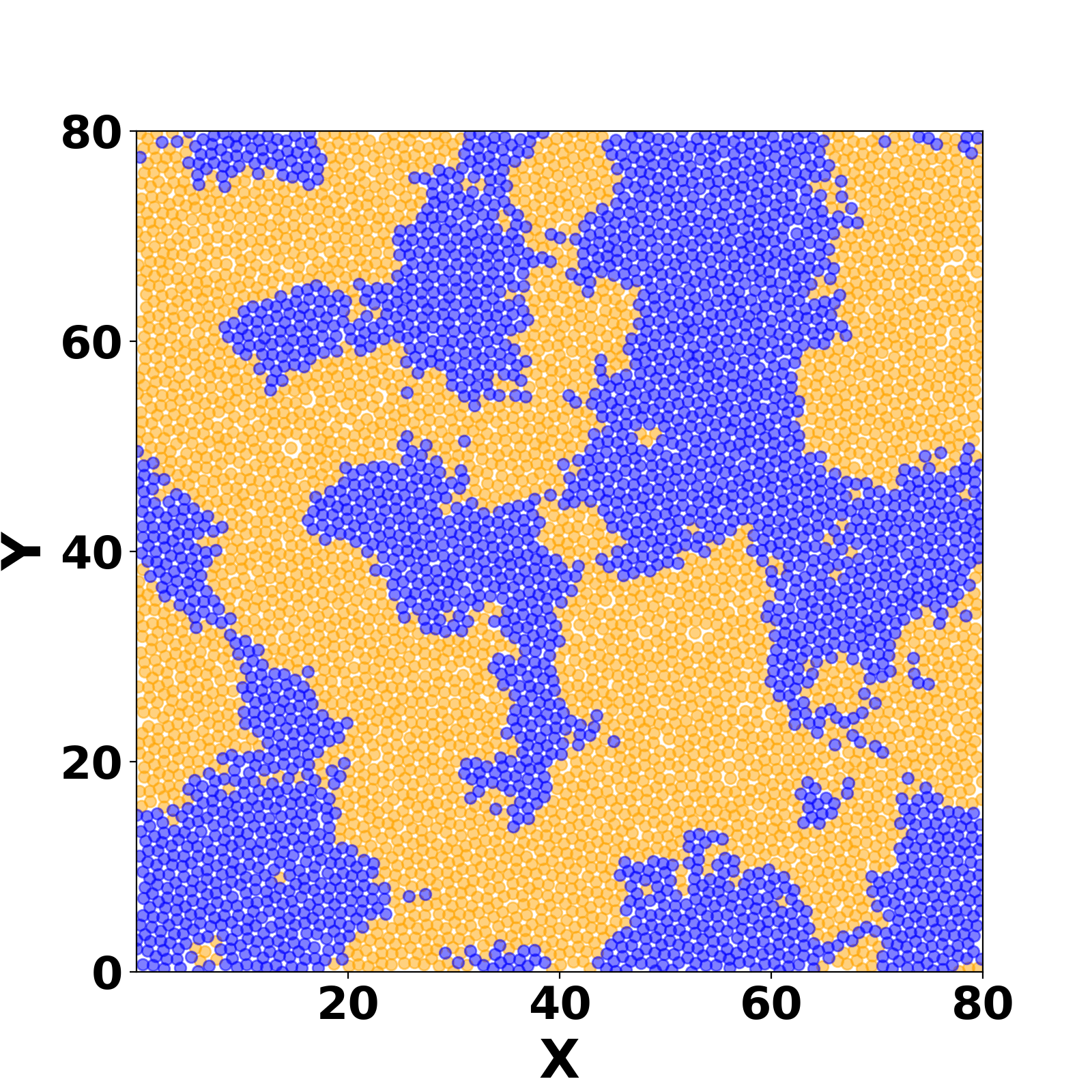}
\caption{$T^*$ = 0.2, Snapshot 3}
   \label{fig:T0.2_20ns}
\end{subfigure}
\hfill
}
%\parbox{3.3in}
%{
\caption{ 2D snapshots of meso-states at different temperatures. The snapshots are took at 10ns lag time.}  
\label{fig:snapshots}
%}
\end{figure}

\begin{figure}
\resizebox{\columnwidth}{!}
{
\begin{subfigure}{0.35\textwidth}
\includegraphics[width=2in]{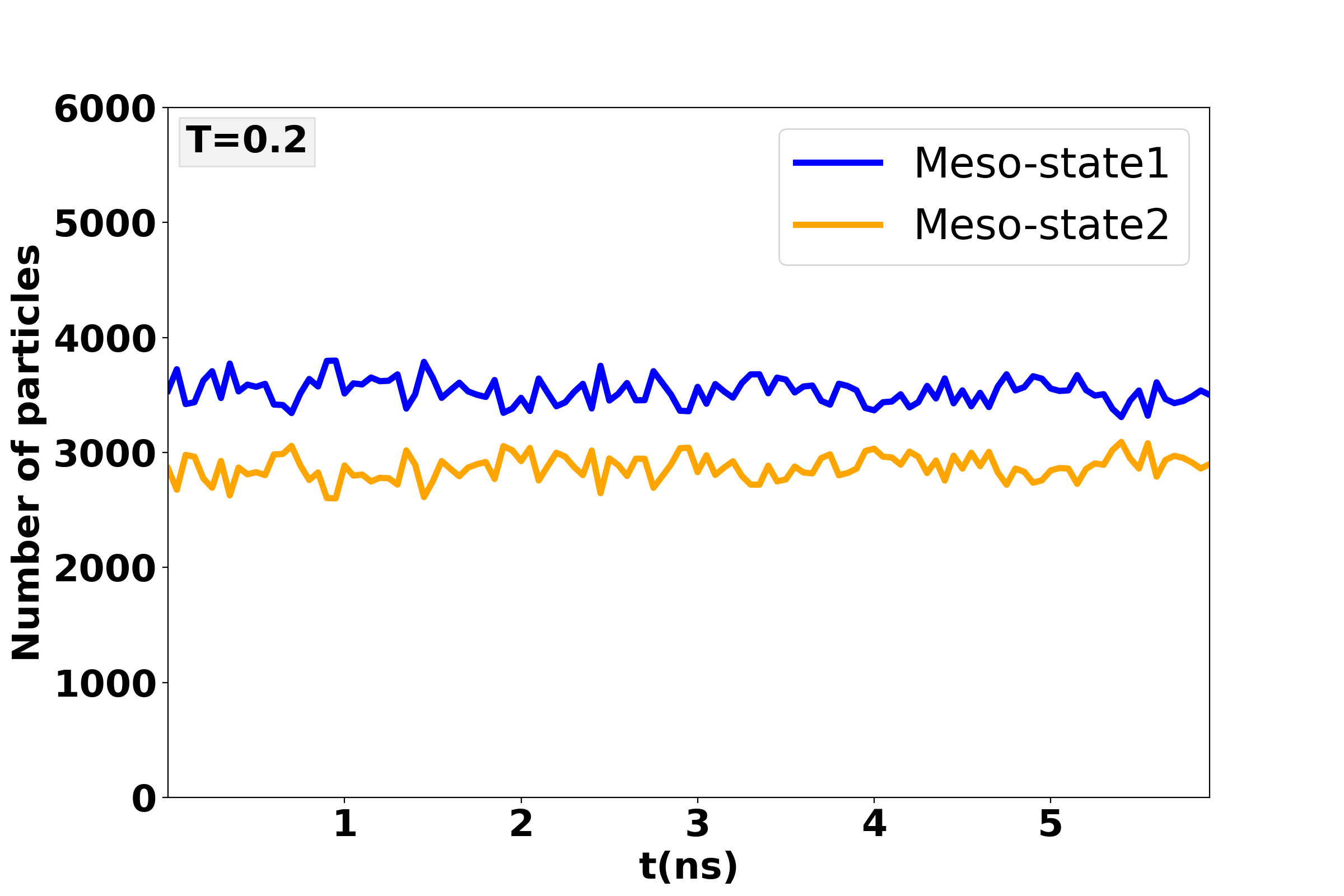}
\caption{$T^*=0.2$}
   \label{fig:fluc_T0.2}
\end{subfigure}
\hfill
\begin{subfigure}{0.35\textwidth}
\includegraphics[width=2in]{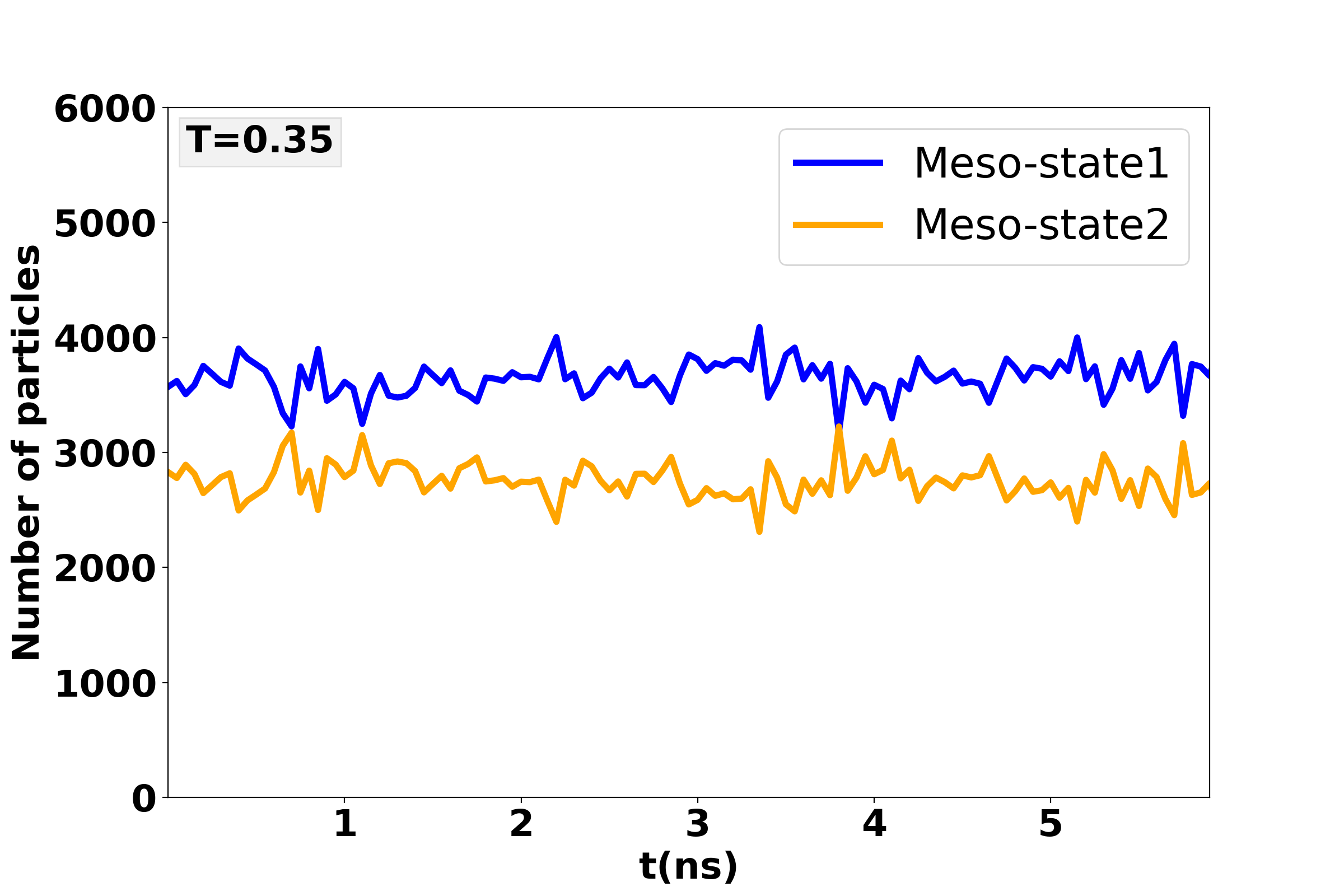}
\caption{$T^*=0.35$}
   \label{fig:fluc_T0.35}
\end{subfigure}
\hfill
}
\resizebox{\columnwidth}{!}
{
\begin{subfigure}{0.35\textwidth}
\includegraphics[width=2in]{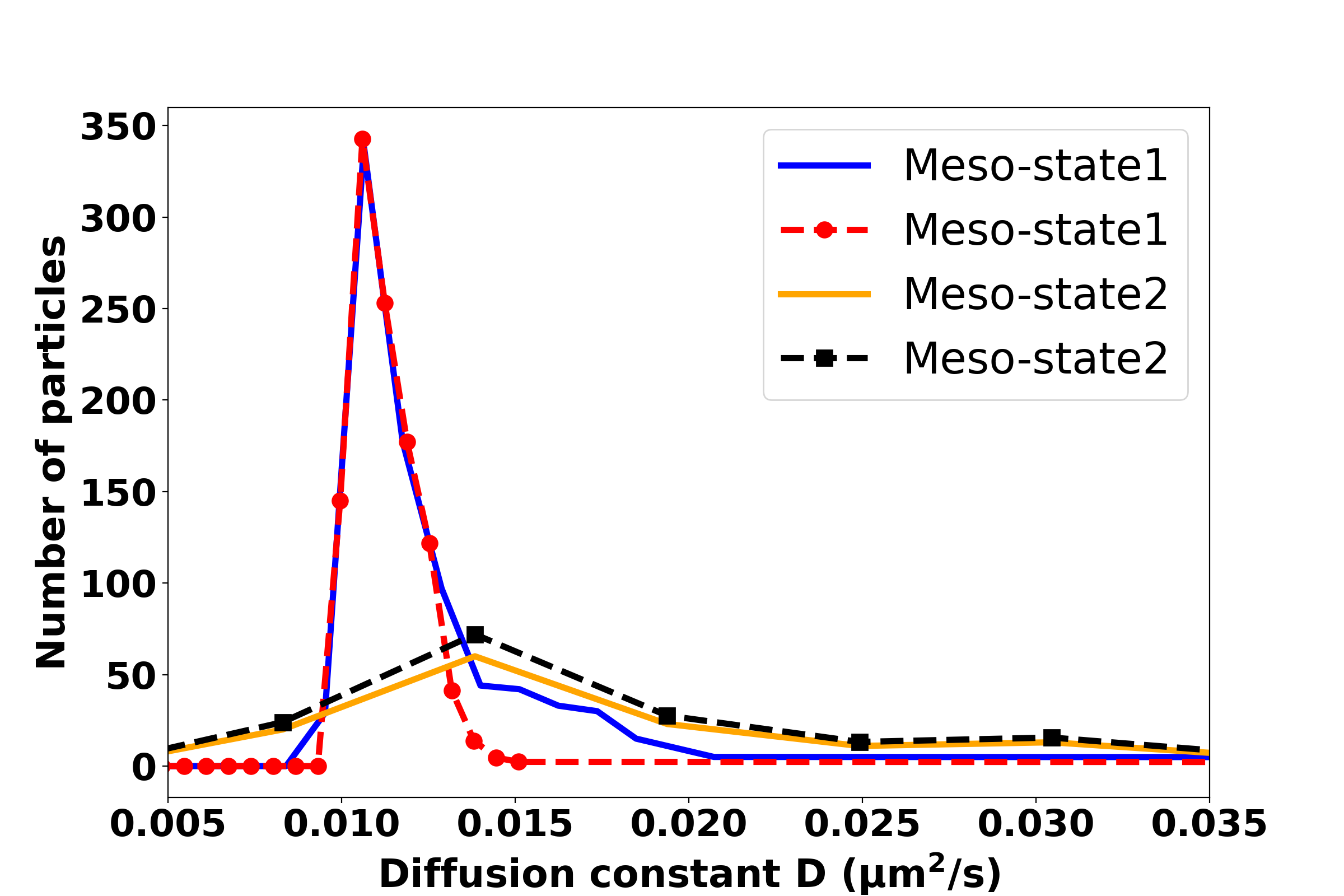}
\caption{D	 distribution}
   \label{fig:D_hist}
\end{subfigure}
\hfill
\begin{subfigure}{0.35\textwidth}
\includegraphics[width=2in]{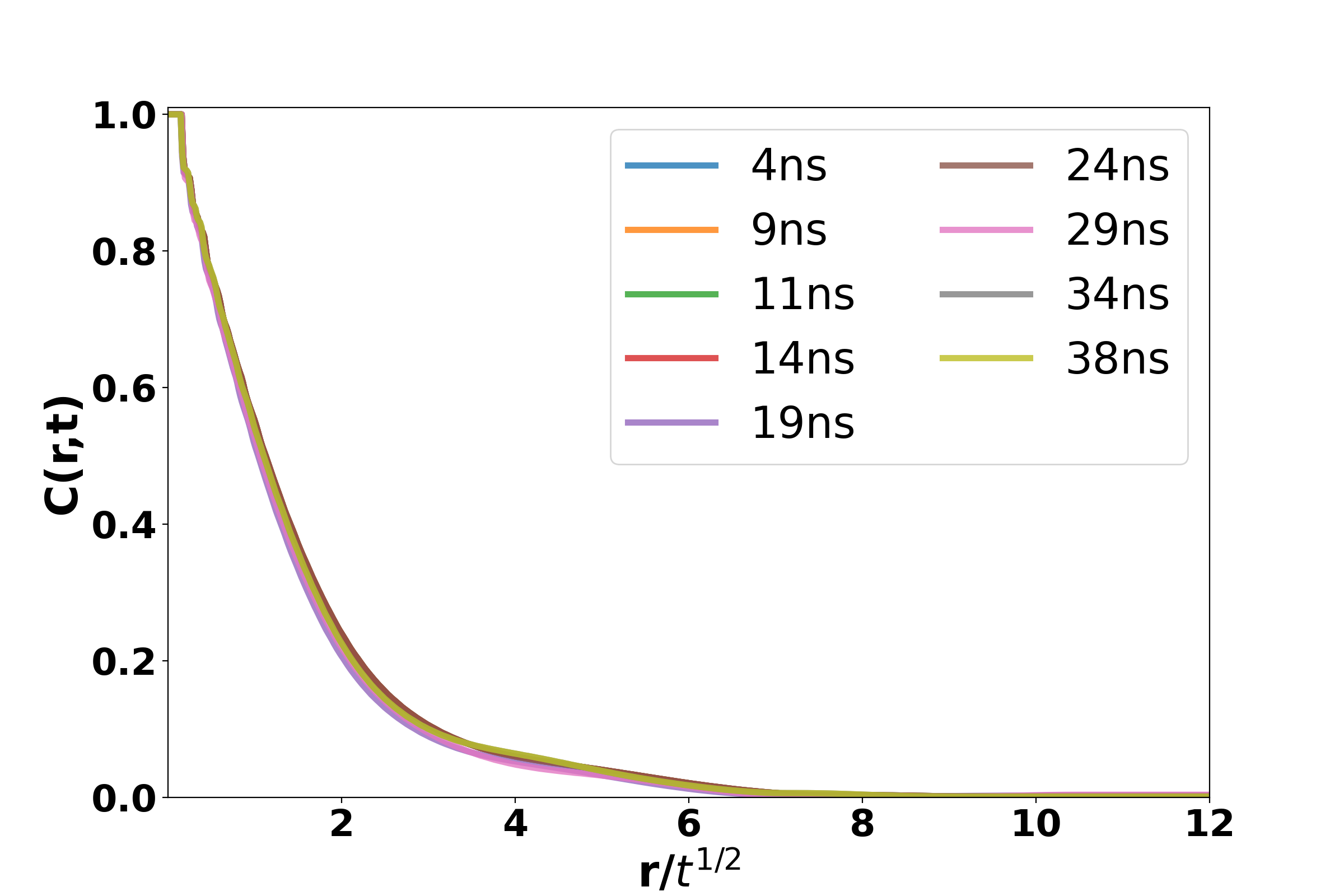}
\caption{$C(r,t)$}
   \label{fig:Crt}
\end{subfigure}
}
%\parbox{3.3in}
%{
\caption{ a-b) Particles fluctuations of meso-states at different temperatures over 6ns.  c) Diffusion constants distribution at $T^*=0.35$. The dash and solid lines represent measurements of diffusion constants distributions for two different domains or patches of a corresponding meso-state.  d) Equal time correlation function $C(r,t)$, where the abscissa is the scaling variable $r / t^{1/2}$ at $T^*$= 0.2.}  
\label{fig:D_Crt}
%}
\end{figure}

Armed by the identities of these domains, heterogeneous dynamics is related to relaxation dynamics of nano-domains which can be measured by their diffusion constant distributions.  Diffusion constants distribution is different for domains belonging to different meso-states while is the same for domains associated with the same meso-state.  After characterizing nano-domains in the configurational space, the core particles, which remain static during the simulation time, are collected in each domain. Those core particles are then used to obtain mean-square displacements to compute diffusion constants through the Einstein relation. Final results  (\ref{fig:D_hist} support the picture that the same diffusion constant distribution of core particles for different domains associated with the same meso-state and different distribution in different meso-states, but the long time scale heterogeneous dynamics reflecting the activated cage escaping need much longer simulations to study.

The stability to form nano-domains is closely related to the onset of cage processes which is signified by the plateau region of diffusion dynamics through MSD \ref{fig:msd}. The plateau physically corresponds to the trapping of particles within the cage. These particles take long times but eventually mange to escape out of the cage and returns to a normal diffusive motion. The timescale of the cage process depends on temperatures, thus a quantitative characterization will need further studies, but some qualitative description can be made by the observation of different configurational snapshots. \ref{fig:snapshots} shows three different  configurations with 10ns lag time at two different temperatures to qualitatively examine the timescale of nano-domains. At $T^*$ = 0.35 which is above the $T^*_g$, \ref{fig:snapshots} a-c show that the particles have sufficient thermal energy to be able to escape out the cage and self-rearrange more freely, thus particles of one domain will likely have chance to hop to other domains whose shapes are frequently changing and become unstable. Meanwhile, the shapes of nano-domains are more static due to a freeze of most of particles at their local cage or domain at $T^*$ = 0.2 (below the $T^*_g$) as illustrated in the \ref{fig:snapshots} d-f. These phenomena are supported by quantitatively assessing particles fluctuation of the domains in \ref{fig:D_Crt} a-b. The particles fluctuate more as the temperature increases because of higher number of mobile particles.

Furthermore, as the system size increases we would expect that a meso-state in the PC-space should be able to bifurcate into more domains or patches in the configurational space to tile up the whole space. \ref{fig:sm4c} and \ref{fig:sm4d} demonstrate our prediction. The meso-state 1 splits into more domains in correspondence to the growing system size, thus the larger the system is, the higher number of patches each meso-state owns.

Finally, to further confirm the nature of these domains reflects a liquid-liquid phase separation after quenching, the kinetics of domain growth following the quenching from a high temperature equilibrium state (normal liquid) state to a deeply supercooled state is calculated.  At the quenched temperature, the morphology of the system evolved either by nucleation or spinodal decomposition with two coexisting thermodynamic states represented by the two meso-states shown in~\ref{fig:d} (we also observed nucleation scenario by change the initial density of the system~\cite{jayme} at the same quenched temperature). It has been known that the non-equilibrium domain growth laws of the system follows a growth law of  characteristic domain size $L(t) \sim t^{1/2}$ for a non-conserved scalar order parameter, which can be measured by an equal-time correlation function $C(r,t)$~\cite{Humayun,Gunton,Mazenko}. The $C(r,t)$ has form:  $C(r,t) = N^{-1}\sum_{i} \psi_i(t)\psi_{i+r}(t)$ where N is the number of particles, $\psi_i$ is the scalar order-parameter which $\psi_{i}$ = +1 for particles identities of state 1 and $\psi_{i}$ = -1 for particles identities of state 2. $i +r$ indicates a neighboring particle displaced by a distance $r$ relative to the reference particle $i$, hence the product of $\psi_i(t)\psi_{i+r}(t)$ will be +1 between pair of particles from the same state and will be -1 otherwise. Since  the domain identities of the particles are known from our classification~\ref{fig:Crt} indeed shows the results of the scaing law  as $L(t) \sim t^{1/2}$  using the autocorrelation function.

\section{Conclusions} \label{sec:conclusion}

The results reported in this study indicate that a general approach can be developed to understand the existence of spatial heterogeneities both dynamically and structurally  in a supercooled liquid. Such heterogeneity results from a liquid-liquid phase separation when the system is cooled rapidly.  In our approach, weighted coordinate numbers (WCNs) are collected at  various frames.  The collection of WCNs this way provides crucial information to avoid statistical homogeneity, hence reveals the unique difference in structure of supercooled liquids in contrast with other works that the structure of supercooled liquids is only investigated by from the features along the {\it g(r)}. These features only  reflect the highly averaged WCNs representation of the system. Local environment of all particles is statistically the same and each particle will have the same CNs which lacks enough details to provide any realistic description of the spatial heterogeneities. 

PCA with K-means and GM clustering provides a robust method of classifying particles into different thermodynamic states where one state forms domains surrounded by another state in the configurational space. Representations of partial rdf, bimodal distribution in $\overline{WCNs}$, density and pressure profiles,  and domain growth scaling law clearly support our physical interpretation of the liquid-liquid phase separation picture. Given the spatial classification scheme, dynamics of such meso-states shows difference in the distribution of diffusion constants, hence the dynamical heterogeneities may be the results of different thermodynamic states after phase separation. Therefore, our classification scheme paves the way for further statistical mechanics analysis of supercooled liquids.

\section{Acknowledgement}

This work is supported by the Division of Chemical and Biological Sciences, Office of Basic Energy Sciences, U.S. Department of Energy, under Contact No. DE-AC02-07CH11358 with Iowa State University. 

\appendix

\section{Bimodality of $\overline{WCNs}$ distribution} \label{sec:appen}
\renewcommand{\thefigure}{A\arabic{figure}}
\setcounter{figure}{0}

\begin{figure}[h!] 
\resizebox{\columnwidth}{!}
{
\begin{subfigure}{0.35\textwidth}
\includegraphics[width=2.1in]{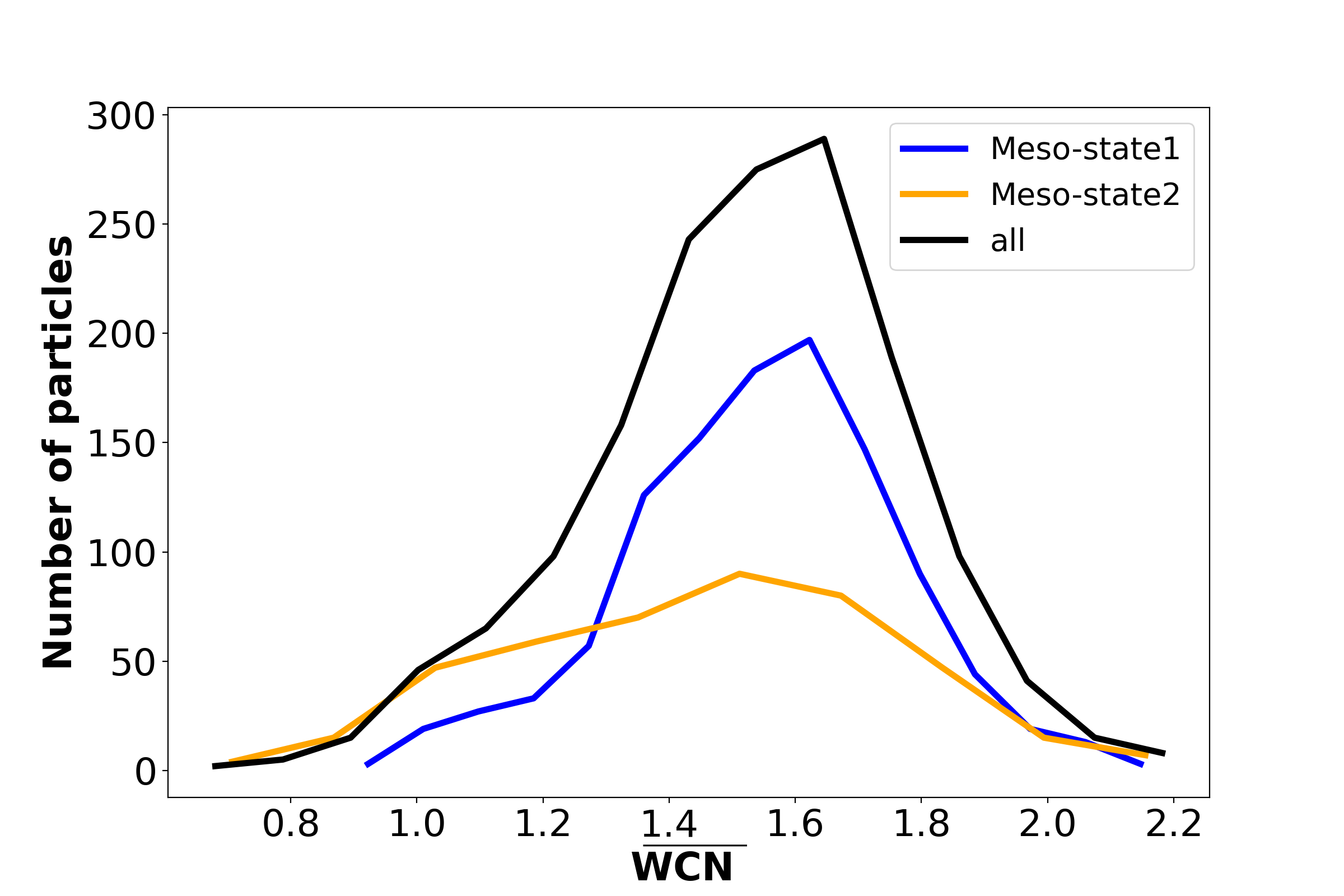}
\caption{ss5}
   \label{fig:s1}
\end{subfigure}
\hfill
\begin{subfigure}{0.35\textwidth}
\includegraphics[width=2.1in]{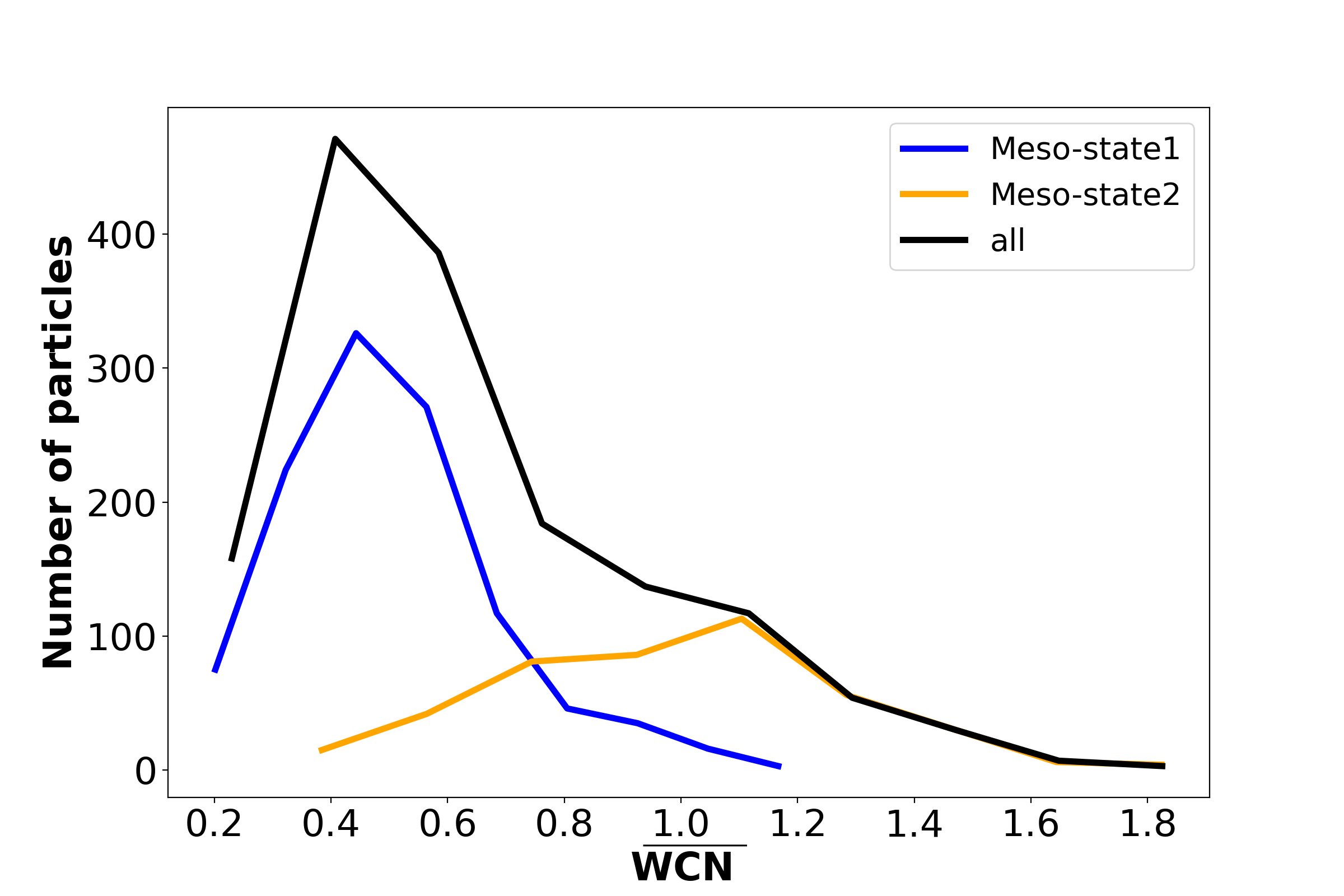}
\caption{ss6}
   \label{fig:s2}
\end{subfigure}
\hfill
}

\resizebox{\columnwidth}{!}
{
\begin{subfigure}{0.35\textwidth}
\includegraphics[width=2.1in]{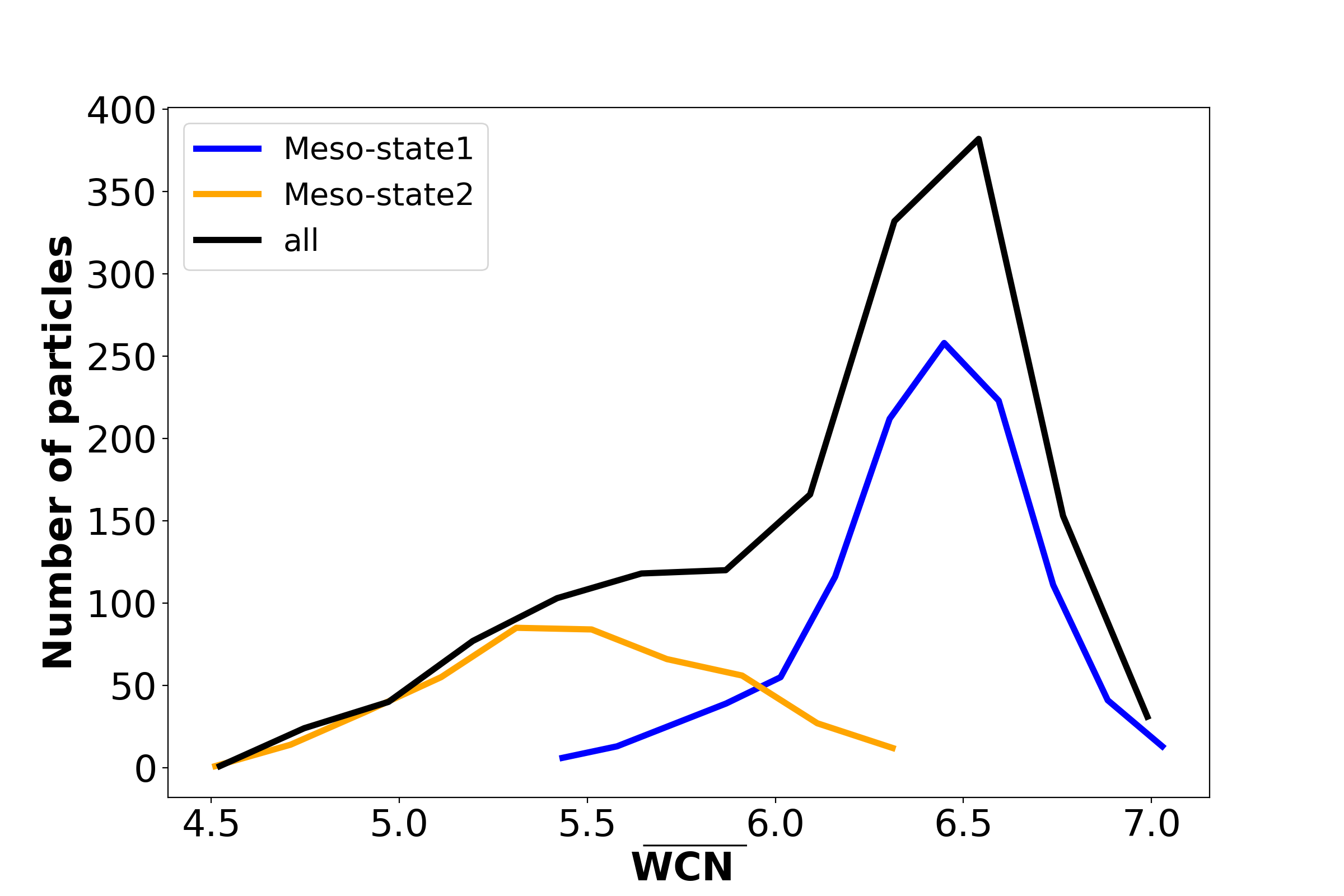}
\caption{ss7}
   \label{fig:s3}
\end{subfigure}
\hfill
\begin{subfigure}{0.35\textwidth}
\includegraphics[width=2.1in]{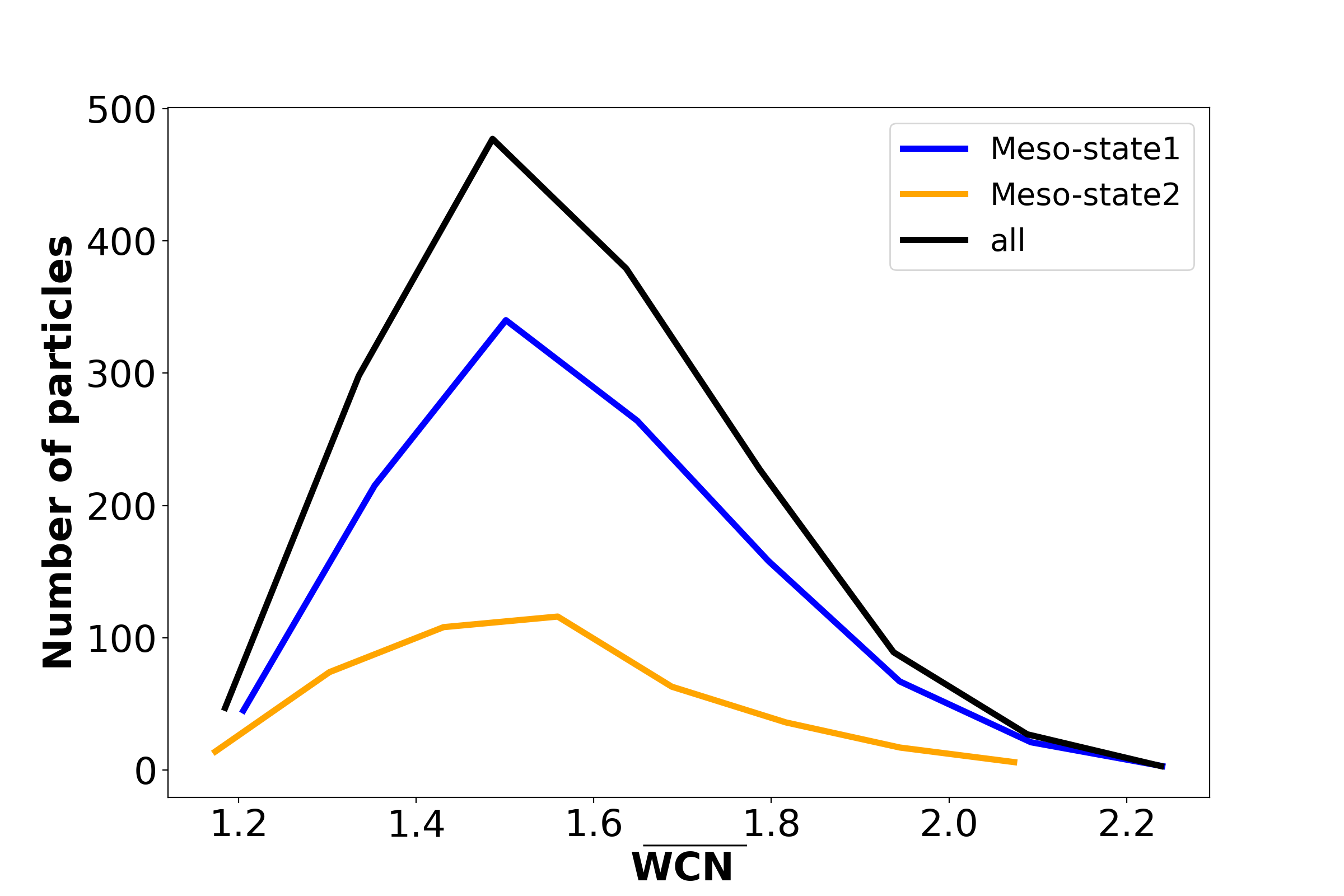}
\caption{ss8}
   \label{fig:s4}
\end{subfigure}
}

\resizebox{\columnwidth}{!}
{
\begin{subfigure}{0.35\textwidth}
\includegraphics[width=2.1in]{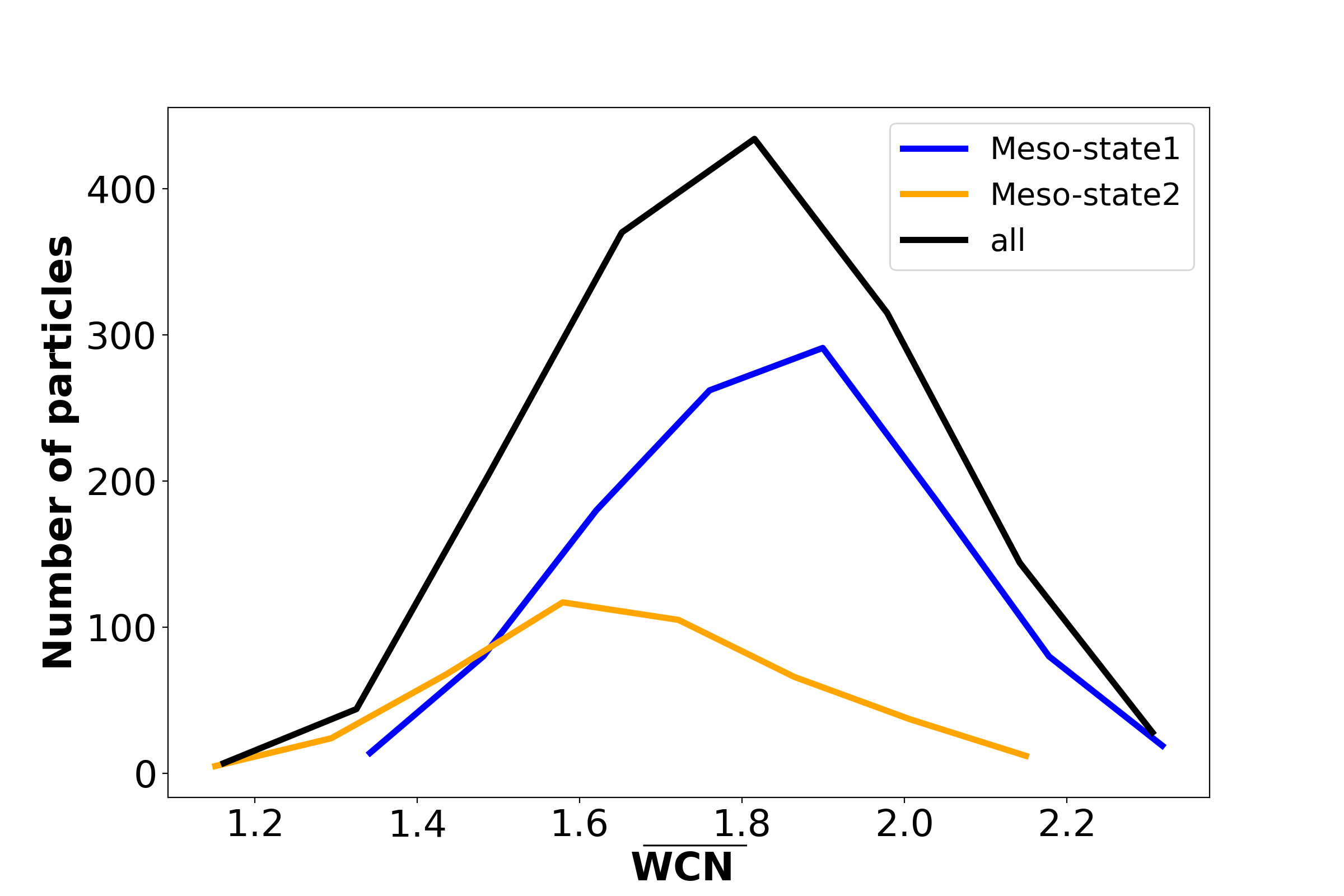}
\caption{ss9}
   \label{fig:s5}
\end{subfigure}
\hfill
\begin{subfigure}{0.35\textwidth}
\includegraphics[width=2.1in]{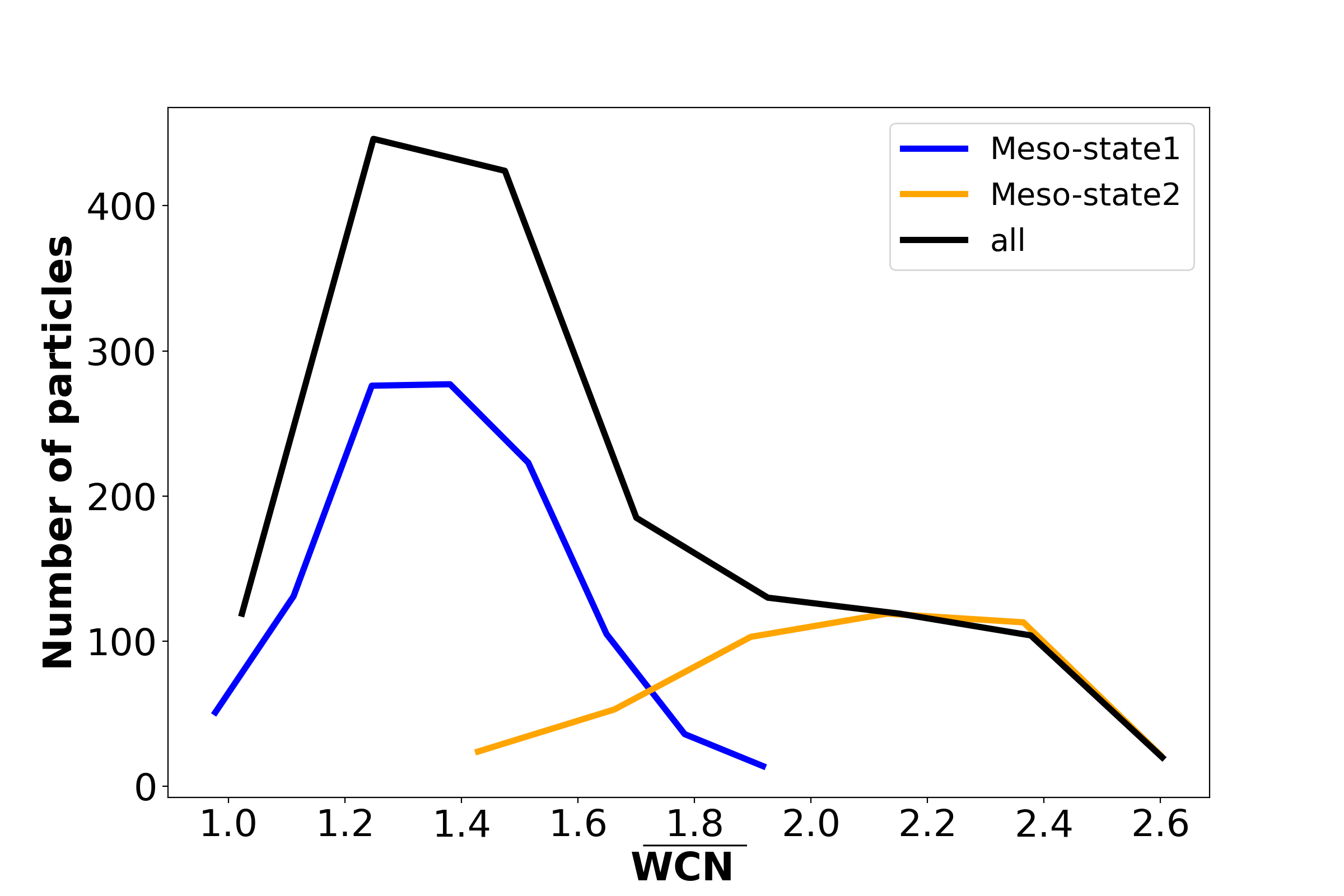}
\caption{ss10}
   \label{fig:s6}
\end{subfigure}
}
\caption{ Weighted $\overline{WCNs}$ distribution for other six solvation shells of each meso-state and whole system at $T^*$ = 0.35, respectively. (ss stands for the solvation shell).}  
\label{fig:wcn_dist_si}
%}
\end{figure}

Bimodality of $\overline{WCNs}$ distribution is described in the main text. \ref{fig:wcn_dist_si} present examples of bimodal distribution along the fifth to tenth solvation shells at $T^* =0.35$.

\bibliographystyle{apsrev4-1}
\bibliography{nano_domain_ljg} 

\end{document}